\documentclass{article}

\usepackage{arxiv}

\usepackage[utf8]{inputenc} % allow utf-8 input
\usepackage[T1]{fontenc}    % use 8-bit T1 fonts
\usepackage{hyperref}       % hyperlinks
\usepackage{url}            % simple URL typesetting
\usepackage{booktabs}       % professional-quality tables
\usepackage{amsfonts}       % blackboard math symbols
\usepackage{nicefrac}       % compact symbols for 1/2, etc.
\usepackage{microtype}      % microtypography
\usepackage{lipsum}		% Can be removed after putting your text content
\usepackage{graphicx}
\usepackage{doi}
\usepackage{wrapfig}
\usepackage{multicol}
\usepackage{multirow}
\usepackage{bm}
\usepackage{xcolor}

\title{Production and Decay of Polarized Hyperon-Antihyperon Pairs}

\author{\href{https://orcid.org/0000-0002-3490-9584}{\includegraphics[scale=0.06]{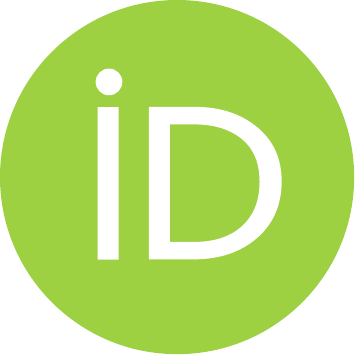}\hspace{1mm}Karin ~Schönning}\thanks{Corresponding author} \\
	Department of Physics and Astronomy\\
	Uppsala University\\
	Uppsala, Sweden \\
	\texttt{karin.schonning@physics.uu.se} \\
\And
	\And
	\href{https://orcid.org/0000-0003-1089-9200}{\includegraphics[scale=0.06]{orcid.pdf}\hspace{1mm}Varvara ~Batozskaya} \\Institute of High Energy Physics \\ Beijing, China \\ and National Centre for Nuclear Research \\ Warsaw, Poland\\
\And
	\href{https://orcid.org/0000-0001-6280-3851}{\includegraphics[scale=0.06]{orcid.pdf}\hspace{1mm}Patrik ~Adlarson} \\Department of Physics and Astronomy\\
	Uppsala University\\
	Uppsala, Sweden \\
\And
\href{https://orcid.org/0000-0002-7671-7644}{\includegraphics[scale=0.06]{orcid.pdf}\hspace{1mm}Xiaorong ~Zhou} \\University of Science and Technology in China \\ Hefei, China\\
}

\renewcommand{\shorttitle}{Hyperon Physics}

\hypersetup{
pdftitle={Production and Decay of Polarized Hyperon-Antihyperon Pairs},
pdfsubject={nucl-exp},
pdfauthor={Karin ~Schönning, Varvara ~Batoszkaya, Patrik Adlarson, Xiaorong Zhou},
pdfkeywords={Hyperon, Polarization, CP violation, Form Factors, BESIII},
}

\begin{document}
\maketitle

\begin{abstract}
	Polarized hyperon-antihyperon pairs shed light on various unresolved puzzles in contemporary physics: How the strong interaction confines quarks into hadrons, how accurately the Standard Model describes microcosmos and even why our universe consists of so much more matter than antimatter. Thanks to their weak, parity violating decays, hyperons reveal their spin properties. This can be exploited \textit{e.g.} the decomposition of the electromagnetic structure of hyperons, precision tests of flavour symmetry and searches for CP violation. At the BESIII experiment at BEPC-II, Beijing, China, hyperon-antihyperon pairs can be produced in abundance. Recently collected large data samples have triggered the development of new methods that provide unprecedented precision and a plethora of new results have emerged. When applied at future high-intensity facilities like PANDA and STCF, precision physics will be taken to a new level which can contribute to the solution to the aforementioned puzzles.
\end{abstract}

% keywords can be removed

\section{Introduction}

Hyperons offer a unique tool to study some of the most challenging problems in contemporary physics. In this review, we will in particular demonstrate how polarized and entangled hyperon-antihyperon pairs can be exploited as a precision instrument to probe various aspects of the Standard Model (SM) of particle physics. Though immensely successful in describing the elementary particles and their fundamental interactions, the SM falls short in describing the properties of complex systems such as nucleons, nuclei or even the evolution and composition of our universe. Many puzzles in modern physics are related to the limitations of the SM and they fall into two main categories: those whose solution should be sought within the SM, and those where one needs to go beyond. 

Within the SM, it remains a challenge to quantitatively describe the main building block of the visible universe: the nucleon. Its mass, size and structure emerge from the complex dynamics of the strong interaction. The theory of Quantum ChromoDynamics (QCD), an inherent part of the SM, describes these interactions from first principles, but its non-Abelian nature makes it difficult to make quantitative predictions at distances where quarks form hadrons. Replacing one or several light quarks in the nucleon with heavier strange or charm, gives deeper insights into these interactions.

Various experimental observations, such as neutrino oscillations, the need for dark matter to explain cosmological observations and the matter-antimatter asymmetry of the universe, raise the question of whether there is a more fundamental theory beyond the SM that encompasses these phenomena. Intensive searches for physics beyond the SM (BSM) are ongoing at the high energy frontier, in a quest to find hitherto unknown, heavy particles manifesting a new fundamental theory. A complementary approach is the precision frontier: new aspects of particles whose existence is established since long. By studying the properties of these particles, \textit{e.g.} measuring quantities that should either be zero, or that can be calculated with immense precision, physics beyond the SM would reveal itself by significant deviations from these predictions.

\begin{wrapfigure}{l}{0.35\textwidth}
\begin{center}
%\vspace{-4mm}
\includegraphics[width=0.35\textwidth]{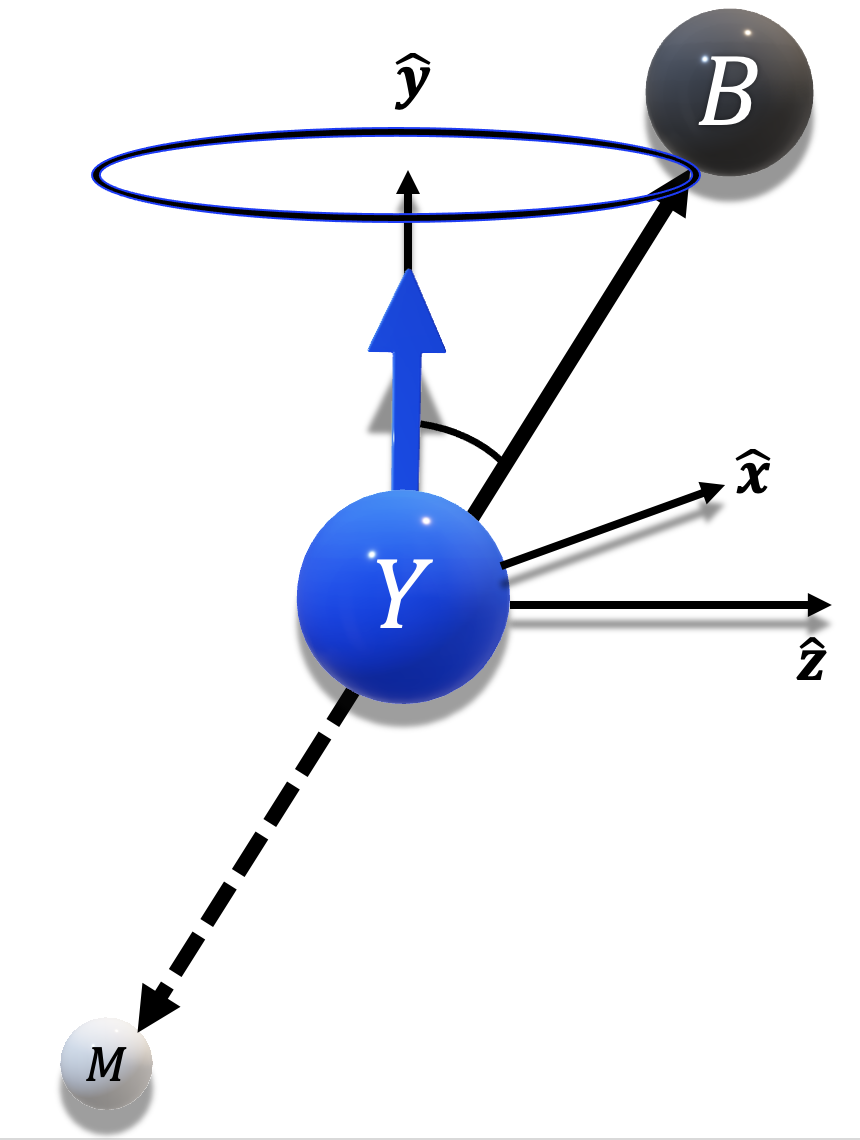}
%\vspace{-7mm}
\end{center}
\caption{The $Y \to BM$ decay, with the spin direction of $Y$ along the $y$-axis.}
\label{fig:hypdecay}
\end{wrapfigure} 

Hyperons are well suited for precision physics through their weak, parity violating decays, giving straightforward experimental access to their spin properties. This is in contrast to \textit{e.g.} protons, for which dedicated polarimeter detectors are required for this purpose. In hyperon decays, the daughter particles are emitted according to the direction of the spin of the mother hyperon. For example, consider a two-body decay $Y \to B M$, where $Y$ is a spin 1/2 hyperon, $B$ a spin 1/2 baryon and $M$ a pseudoscalar meson as illustrated in Fig. \ref{fig:hypdecay}. The angular distribution of $B$ in the rest system of $Y$ with respect to some reference axis $\vec y$ is given by \cite{ParticleDataGroup:2022pth,Bigi:2000yz}
\begin{equation}
W(\cos\theta_B)=\frac{1}{4\pi}(1+\alpha P_y \cos\theta_B).
\label{eq:decay}
\end{equation}

\noindent The polarization $P_y = P_y(\cos\theta_{Y})$ carries information about the production mechanism, while the decay asymmetry $\alpha$ only depends on the decay process. The parameter $\alpha$ is proportional to the real part of the product of the interfering parity violating and parity conserving decay amplitudes \cite{Lee:1957qs}. The imaginary part is accessible in sequential hyperon decays, \textit{i.e.} hyperons that decay weakly into other hyperons \cite{Faldt:2017yqt,erik}.  
Eq. \ref{eq:decay} demonstrates an example of how physical parameters, such as $\alpha$ and $P_y$, can be retrieved from measurable quantities such as $\cos\theta_B$ and $\cos\theta_Y$. This feature is crucial in the work presented in this review.

\section{Fundamental questions}
\label{sec:questions}

\subsection{Structure at the femtometer scale}

 The nature of the strong interaction manifests in the properties of the nucleon, such as its mass and size. These are connected since a mass-scale is a prerequisite to confine colour charges into colour singlets of finite size \cite{Binosi:2022djx}. The structure, \textit{i.e.} the distribution and motion of the quarks inside hadrons, is crucial to understand the underlying mechanisms. Recently, tremendous efforts in atomic and subatomic physics as well as hadron theory have led to rapid progress in solving the so-called proton radius puzzle \cite{Gao:2021sml}. A coherent picture of the emergence of hadron properties requires corresponding studies of similar but different quark systems, such as neutrons \cite{Atac:2021wqj, BESIII:2021tbq} and hyperons. The latter contains one or several heavy quarks such as strange and charm, which should have an impact on their inner structure, including properties like the charge radius. Measured radii of pions \cite{SELEX:2001fbx} and kaons \cite{AMENDOLIA1986435} indicate that this is indeed the case.

 Electromagnetic form factors (EMFFs) are structure functions that have the advantage that they are experimentally accessible for both protons, neutrons and hyperons. \textit{Space-like} electric $G_E$ and magnetic $G_M$ form factors, probed in elastic electron-baryon ($e^-B \to e^-B$ ) scattering, are related to the charge- and magnetisation densities, respectively \cite{Punjabi:2015bba}. However, since the hyperons are unstable, they are unfeasible as beams or targets and as a consequence, their space-like EMFFs are hard to access experimentally. Instead, the \textit{time-like} EMFFs constitute the most viable structure observables for hyperons \cite{Pacetti:2014jai}. These can be probed in either $e^+e^- \to \gamma^* \to Y_1 \, \bar Y_2$ reactions or in so-called Dalitz decays $Y_1 \to Y_2 \gamma^*, \gamma^* \to e^+e^-$. The experimentally accessible time-like and the intuitive space-like EMFFs are related \textit{via} dispersion relations \cite{Belushkin:2006qa}.  

Space-like EMFFs are real functions of $q^2$, whereas the time-like ones are complex. The electric and the magnetic form factor of a spin 1/2 hyperon have a relative phase $\Delta\Phi$ \cite{Dubnickova:1992ii}, reflecting fluctuations of the $\gamma^*$ into \textit{e.g.} a $\pi\pi$ intermediate state.  At a certain scale $q^2_{asy}$, the time-like and the space-like EMFFs should converge to the same real value, \textit{i.e.} their phase will approach an integer multiple of $\pi$ \cite{Matveev:1973uz,Brodsky:1973kr}. From the asymptotic behaviour of the EMFF phase, we can even extract information about space-like quantities such as the charge radius \cite{Mangoni:2021qmd}. But how do we measure this phase?

In the $e^+e^- \to Y\bar{Y}$ reaction, a non-zero phase manifests in a polarized final state, even if the colliding beams are unpolarized \cite{Dubnickova:1992ii}. The polarization has a well-defined dependence on the hyperon scattering angle, the form factor ratio $R = |G_E/G_M|$ and the phase $\sin\Delta\Phi$ \cite{Faldt:2017kgy}. Therefore, the phase can be calculated from the polarization, which in turn can be extracted from the angular distribution of the decay products, according to Eq.~\ref{eq:decay}. 

In the vicinity of vector charmonium such as $J/\Psi$, $\Psi(3686)$ and $\Psi(3770)$, the production of most hyperon-antihyperon pairs is strongly influenced by processes involving these vector states. In some cases, in particular for $J/\Psi \to Y\bar{Y}$, this process dominates. For others, such as the isospin-breaking $J/\Psi \to \Lambda\bar{\Sigma}^0$, it is suppressed. In the former cases, the form factors describe the vector decay and are denoted \textit{psionic} form factors, in contrast to electromagnetic form factors \cite{Faldt:2017kgy}. Though psionic form factors are straightforward to measure with high precision, their physical meaning is yet to be understood.

\subsection{Searching Beyond the Standard Model}

\subsubsection{CP violation}
 Despite its great success, the SM does not provide an explanation for the matter-antimatter asymmetry of the universe; Why does our universe consists of matter, and not antimatter? Unless fine-tuned in the Big Bang, the matter abundance must have been generated dynamically. The mechanism proposed by A.~Sakharov, \textit{Baryogenesis} \cite{Sakharov:1967dj}, has become the most long-standing candidate for such a process. However, several criteria need to be fulfilled for this to be possible: i) the existence of processes that violate baryon number, ii) the existence of processes that violate the conservation of charge conjugation (C) and the combined charge conjugation and parity (CP), and iii) the processes i) and ii) must occur outside thermal equilibrium. Hyperons provide a sensitive instrument to test criterion ii) as we will demonstrate in this report. 
 
CP violation in weak interactions is accommodated in the SM through the Cabibbo-Kobayasi-Maskawa (CKM) mechanism \cite{Cabibbo:1963yz,Kobayashi:1973fv}. However, the predicted CP violations are not sufficient to explain the observed matter abundance of the universe. Hence, the search for CP violation is connected to the search for physics beyond the SM. Decays of entangled hyperon-antihyperon pairs provide a unique hunting-ground for such signals. Hyperons decay through a complex interplay of strong, CP conserving interactions and weak, possibly CP violating ones and the challenge is to interpret the decay patterns of hyperons and antihyperons in terms of CP violating amplitudes. 

While predicting small CP violations in weak interactions, the SM should allow for CP violating strong interactions. This would for instance manifest in a non-zero electric dipole moment (EDM) of the neutron \cite{PhysRevLett.63.589}. However, the upper limit of the neutron EDM has been found to be extremely small \cite{2021135993}, suggesting an unnaturally small CP violation in the strong interaction. The fact that the SM does not explain why this is so small, is referred to as the \textit{strong CP problem} \cite{Peccei:2006as}. Flavour conserving radiative hyperon decays are related to the neutron EDM through flavour symmetry and can therefore constitute a potential test of the strong CP problem \cite{nair}. In addition, they offer a probe for tests of C conservation. Flavour conserving radiative decays of hyperons involve both weak, electromagnetic and strong interactions and probe P and CP symmetry \cite{Hara:1964zz}.

\subsubsection{Flavour Symmetry}

In addition to the aforementioned CP tests, hyperon decays offer independent ways to probe the SM at the precision frontier. In particular, the CKM matrix elements, characterizing the quark flavour mixing, are crucial parameters of the SM. Considering the three lightest quarks up ($u$), down ($d$) and strange ($s$), the matrix elements $|V_{ud}|$ and $|V_{us}|$ can be precisely predicted and also tested independently in semi-leptonic decays ($H_1 \to H_2 l \bar{\nu}_l$) of mesons as well as hyperons. In fact, the interplay between the weak interaction and the hadronic structure is revealed with greater richness in hyperon decays compared to meson dittos. This is thanks to the presence of three valence quarks in the former case, compared to the simpler quark-antiquark structure~\cite{Cabibbo:1963yz}. 

A peculiar feature of the SM is that the electroweak gauge bosons, \textit{i.e.} the $\gamma$, the $Z$ and the $W^{\pm}$, couple in the same way to all leptons, irrespective of generation. This feature, referred to as lepton universality, can be probed in semi-leptonic decays of hyperons~\cite{Chang:2014iba}. 

\section{The BESIII Experiment}

The BEijing Spectrometer (BESIII) \cite{BESIII:2009fln} at the Beijing Electron Positron Collider (BEPC-II) is a multi-purpose experiment that offers unique opportunities to explore strange and single-charm hyperons. The experiment operates at CMS energies ranging from 2.0 GeV up to 4.95 GeV. Hyperon-antihyperon pairs can be produced either in one-photon exchange processes $e^+e^- \to \gamma^* \to Y\bar{Y}$ by off-resonance energy scans, or from vector charmonium decays, \textit{e.g.} $e^+e^- \to J/\Psi \to Y\bar{Y}$.

Since the operation commenced in 2009, BESIII has collected more than 40~fb$^{-1}$ of data, comprising several world-leading data samples including $10^{10}$ $J/\psi$ events \cite{BESIII:2021cxx} and $2.7\cdot10^{9}$ $\psi(3686)$ events. These samples give unprecedented access to the hyperon pairs via $\psi\to Y\bar{Y}$ decays. In addition, fine energy scans have been performed in a range covering the production threshold regions of all strange ground-state hyperons. Recently, more than 3~fb$^{-1}$ of data have been collected above the $\Lambda_{c}^+\bar{\Lambda_{c}}^-$ threshold.

 The BESIII detector covers 93\% of the 4$\pi$ solid angle. A small-cell, helium-based main drift chamber (MDC) surrounding the $e^+e^-$ collision point provides precise tracking of charged particles. The BESIII detector also comprises a time-of-flight system (TOF) based on plastic scintillators, an electromagnetic calorimeter (EMC) made of CsI(Tl) crystals, a muon counter (MUC) made of resistive plate chambers, and a superconducting solenoid magnet with a central field of 1.0 Tesla.

The work presented here is based on a low-energy scan from 2015, including a large sample at 2.396 GeV, a high-energy scan from the $\Lambda_c^+\bar{\Lambda}_c^-$ threshold near 4.6 GeV, collected during 2014, a scan above 4.6 GeV during 2019 - 2021, a sample of $\Psi(3686)$ events collected during 2009 and 2012, $1.2\cdot10^{9}$ $J/\Psi$ events collected in 2009 and 2012, and $10^{10}$ $J/\Psi$ from 2018 and 2019.

\section{Polarized and Entangled Hyperons}
\label{sec:polent}

\subsection{Hyperon-Antihyperon Pair Production}

\begin{figure}[h]
    \centering
    \includegraphics[width=.85\textwidth]{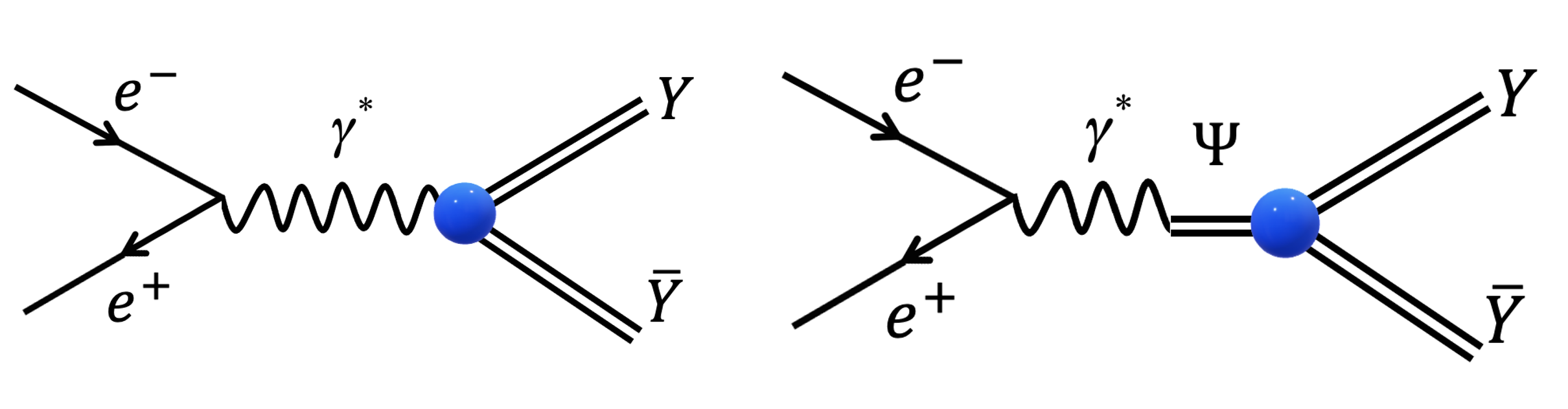}
    \caption{Illustration of the process $ e^+ e^- \to \gamma^* \to Y \overline{Y}$, where $\gamma^*$ denotes a virtual photon (left) and the $ e^+ e^- \to \Psi \to Y \overline{Y}$, where $\Psi$ a vector charmonium resonance (right).}
    \label{fig:coord}
\end{figure}

Consider the production of a pair of a spin 1/2 hyperon and antihyperon in either the $e^+e^- \to \gamma^* \to Y\bar{Y}$ (Fig.~\ref{fig:coord}, left panel) or $e^+e^- \to \Psi \to Y\bar{Y}$ (Fig.~\ref{fig:coord}, right panel), where $\Psi$ is any vector charmonium resonance: $J/\Psi$, $\Psi(3686)$, $\Psi(3770)$ etc. In $e^+e^-$ annihilations, processes with spin-parity $J^P = 1^-$ dominate completely which means that the spin-parity of the final state is considered to be known. In the case of an intermediate virtual photon $\gamma^*$, the process can be parameterised in terms of two complex electromagnetic form factors $G_E(q^2)$ and $G_M(q^2)$ \cite{Dubnickova:1992ii,Gakh:2005hh,Czyz:2007wi,Faldt:2016qee}. The form factors are functions of the momentum transfer $q^2$ carried by the virtual photon. The Born cross section $\sigma_B$ is related to the effective form factor $G_{\rm eff}(q^2)$, a linear combination of $G_E(q^2)$ and $G_M(q^2)$:
\begin{equation}
  G_{\rm eff}(q^2) = \sqrt{\frac{2\tau|G_M(q^2)|^2+|G_E(q^2)|^2}{1+2\tau}}=\sqrt{\frac{\sigma_B \cdot 3q^2}{1+\frac{1}{2\tau}(4\pi\alpha^2\beta)}}
\end{equation}
where $\tau=q^2/(4M_Y^2)$, $\alpha$ is the electromagnetic fine structure constant and $\beta$ is the velocity.

The modulus of their ratio $R=|G_E(q^2)/G_M(q^2)|$ governs the scattering angle $\theta$ between the produced hyperon and the positron beam:
\begin{equation}
\label{eq:scatR}
    \frac{d\sigma_B}{d\Omega}=\frac{2\pi\alpha^2\beta}{4q^2}(1+\cos^2\theta+\frac{R^2}{\tau}\sin^2\theta) = \frac{2\pi\alpha^2\beta}{4q^2}(1+\eta\cos^2\theta),
\end{equation}

where $\eta = (\tau-R^2)/(\tau+R^2)$. The relative phase $\Delta \Phi$ of $G_E(q^2)$ and $G_M(q^2)$ governs the vector polarization and tensor polarization (\textit{i.e.} spin correlations) of the produced hyperon-antihyperon pair \cite{Faldt:2017yqt}. %We define a right-handed reference system with the $z$-axis in the direction of the positron beam and the $y$-axis along the normal of the plane spanned by the positron beam and the outgoing hyperon. 
If the beams are unpolarized, then a vector polarization of the final state hyperons is only allowed in the direction normal to the plane spanned by the incoming beam and the outgoing hyperon:
\begin{equation}
P = \frac{\sqrt{(1-\eta^2)}\sin\theta\cos\theta}{1+\eta\cos^2\theta}\sin\Delta\Phi.
\end{equation}

In addition, there are five non-zero spin correlations. In the case of an intermediate vector charmonium, the electromagnetic form factors are replaced by the so-called psionic form factors $G_{E}^{\Psi}$ and $G_M^{\Psi}$ \cite{Faldt:2017yqt}. The formalism is the same as for an intermediate $\gamma^*$, but by convention, the $\eta$ in Eq. \ref{eq:scatR} is denoted $\alpha_{\Psi}$ and the phase $\Delta\Phi_{\Psi}$. 

If a spin 1/2-spin 3/2 hyperon-antihyperon pair is produced in an initial $J^P = 1^-$ state, an additional Coulomb quadrupole form factor $G_C(q^2)$ \cite{JONES19731} enters the parameterization. An equivalent description is to use three helicity parameters as shown in Ref. \cite{Perotti:2018wxm}. In a similar way, the production of a hyperon-antihyperon pair which are both spin 3/2 can be parameterized by $G_E(q^2)$, $G_M(q^2)$, $G_C(q^2)$ and the octupole form factor $G_O(q^2)$, or equivalently by four helicity amplitudes \cite{Perotti:2018wxm, Kopliovich}.

\subsection{Hadronic Hyperon Decays}

\begin{figure}
    \centering
    \includegraphics[width=.95\textwidth]{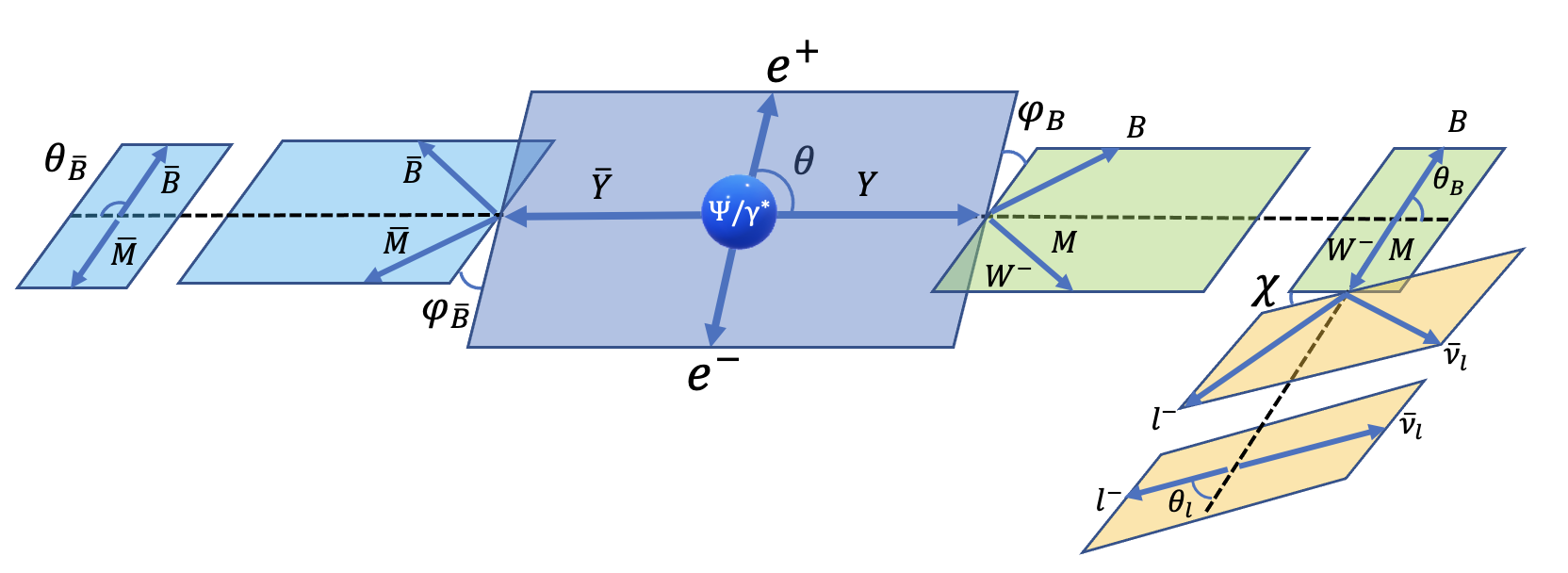}
    \caption{The reaction $ e^+ e^- \to Y \overline{Y}$ which decays either hadronically ($B M$, to the left) or semi-leptonically ($B W^{-}$, to the right). Theplanes illustrate different reference frames, where those where particles are emitted back-to-back represent the rest frame, or the helicity frame, of the decaying particle.  %\textcolor{blue}{Make general / consistent with figure 1 (p -> B, pi -> M)}
    }
    \label{fig:decayplanes}
\end{figure}

The self-analysing hyperon decays, mentioned in the introduction, provide a powerful diagnostic tool to study 
\begin{itemize}
    \item Structure functions, such as electromagnetic form factors.
    \item The spin of the mother hyperon.
    \item Parity violating and parity conserving decay amplitudes.
\end{itemize}

\noindent In the following, we will look into two-body hadronic decays of spin 1/2 or spin 3/2 hyperons decaying into a baryon and a pseudoscalar mesons. 

\subsubsection{Direct Decays}
\label{sec:direct}
By direct decay, we consider the case when a hyperon $Y$ either decays into a stable baryon $B$ and a pseudoscalar meson $M$, or the case when the hyperon decays into another hyperon but where the subsequent decay of the daughter hyperon is not measured or where the daughter is a stable baryon. This process is schematically shown in Figure \ref{fig:decayplanes}. The production plane of the $e^+e^- \to Y\overline{Y}$ process is shown in the middle and the antihyperon decaying into a two-body hadronic state is shown to the left; to the mid-left in the centre-of-mass of the reaction and to the very left in the rest system of the decaying antihyperon. 

\begin{figure}
    \centering
    \includegraphics[width=.95\textwidth]{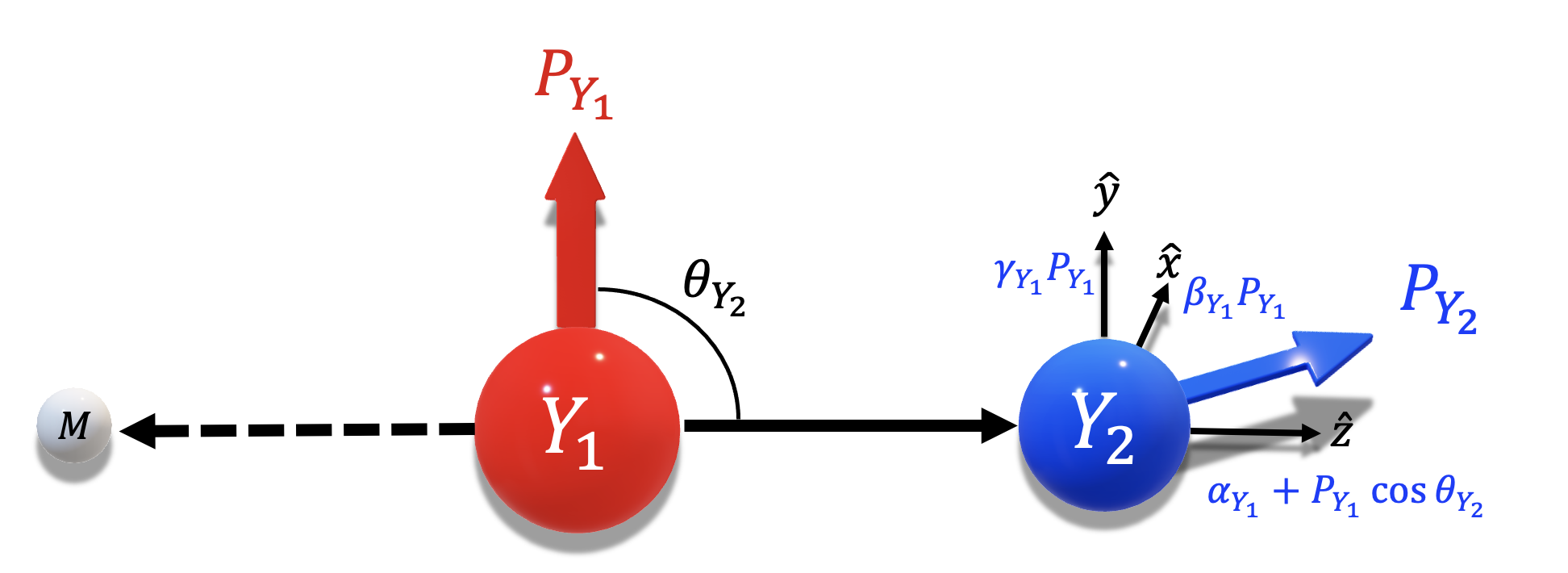}
    \caption{The daughter hyperon $Y_2$ receives longitudinal [$\alpha_{Y_1} + P_{Y_{1}}\cos\theta_{Y_2}$]$\hat{\bf{z}}$ and transversal polarization components $\beta_{Y_1}\hat{\bf{x}}$ and $\gamma_{Y_1}\hat{\bf{y}}$ from the mother hyperon $Y_1$. }
    \label{fig:alphabetaphi}
\end{figure}

The decay of a spin $1/2$ baryon is described by a parity conserving ($P$-wave) and a parity violating ($S$-wave) amplitude. A spin 3/2 hyperon decay can be described in a similar way, but with the $S$-wave replaced by a $D$-wave. The real part of their products is the aforementioned $\alpha_Y$ decay parameter from Eq. \ref{eq:decay}, while the imaginary part is denoted $\beta_Y$ \cite{Lee:1957qs}. A third parameter, $\gamma_Y$, is defined by the difference of the squares of the parity conserving and parity violating amplitudes. These three so-called Lee-Yang parameters fulfil $\alpha_Y^2 + \beta_Y^2 + \gamma_Y^2 = 1$. Furthermore, a parameter $\phi_Y$ can be defined according to
\begin{equation}
    \beta_Y = \sqrt{1-\alpha_Y^2}\sin\phi_Y, ~~~\gamma_Y = \sqrt{1-\alpha_Y^2}\cos\phi_Y.
    \label{eq:leeyang}
\end{equation}
Hence, the decay is completely described by two independent parameters $\alpha_Y$ and either $\beta_Y$, $\gamma_Y$ or $\phi_Y$. The parameter $\alpha_Y$ can be understood more intuitively as the longitudinal polarization induced by the decay, as illustrated in Figure \ref{fig:alphabetaphi}. The longitudinal polarization of the daughter will hence be $\alpha_Y$ plus the transferred polarization from the mother. The latter depends on the decay angle of the daughter. Similarly, $\beta_Y$ and $\gamma_Y$ quantify the transversal polarization transfer.

The formalism of the full process $e^+e^- \to \gamma^{*}/\Psi \to Y\overline{Y}, Y \to BM, \overline{Y} \to \overline{BM}$ has been outlined in Refs. \cite{Faldt:2017kgy,Faldt:2016qee} and demonstrates how the form factors and the decay asymmetry parameters $\alpha_Y$ and $\alpha_{\bar{Y}}$ can be accessed from the joint angular distribution of the decay products $B, M, \overline{B}$ and $\overline{M}$. By so-called single-tag analysis, where either the hyperon or the antihyperon decay is measured while the undetected part is reconstructed from missing kinematics, the quantity $\sin\Delta\Phi$ can be measured \cite{Faldt:2016qee, Thoren:2022exe}. Although the phase $\Delta\Phi$ can only be determined with a $\Delta\Phi/180^{\circ}-\Delta\Phi$ ambiguity, it still provides useful information about the structure \cite{Mangoni:2021qmd}.

From the decay asymmetry parameters, we can construct a CP test: Exact CP symmetry implies equal decay patterns for particles and antiparticles, \textit{i.e.} $\alpha_{Y} = -\alpha_{\bar{Y}}$. Hence, the asymmetry
\begin{equation}
\label{eq:acp}
    A_{CP} = \frac{\alpha_Y+\alpha_{\bar{Y}}}{\alpha_Y-\alpha_{\bar{Y}}}
\end{equation}
quantifies the degree of CP violation. The asymmetry $A_{CP}$ is straightforward to access for all weakly decaying ground-state hyperons. Strange hyperons have the advantage that they decay with an appreciable fraction into two-body hadronic states, hence it is feasible to collect large exclusive $Y\overline{Y}$ data samples where the decays of the hyperon and the antihyperon can be studied simultaneously.

The disadvantage of the $A_{CP}$ is that its sensitivity to CP violation is somewhat limited. This is because hadronic decays involve strong, CP conserving amplitudes as well as weak, possibly CP violating, amplitudes. For spin 1/2 hyperon decays, the former are quantified by strong phase shifts $\delta_S$ and $\delta_P$ in the final state $BM$ interaction. Here, $S$ and $P$ denote $S$-waves and $P$-waves, respectively. The strangeness ($|\Delta S|=1$) and isospin ($\Delta I = 1/2$) changing transition in the $Y \to BM$ decay induce weak, CP violating phases $\xi_S$ and $\xi_P$. Hence, the phase difference $\xi_P- \xi_S$ arises from the interference between $S$-waves and $P$-waves and if different from zero, it indicates CP violation. This is similar to kaon decays, where CP violation is induced by an interference between the isospin transition amplitudes $\Delta I = 1/2$ and $\Delta I = 3/2$. Hyperon decays also involve both transitions but the $\Delta I = 1/2$ is believed to dominate \cite{Donoghue:1986hh}. Assuming this is the case, the weak, CP violating phase $(\xi_P- \xi_S)$ can be extracted from the measurable asymmetry $A_{CP}$ by~\cite{Donoghue:1985ww,Donoghue:1986hh,Tandean:2002vy}
\begin{equation}
\label{eq:acpphase}
    A_{CP} \approx -\tan(\delta_P-\delta_S)\tan(\xi_P-\xi_s).
\end{equation}
However, we see that if there is no $BM$ final state interaction, \textit{i.e.} if the difference $(\delta_P - \delta_S)$ is close to zero, the $A_{CP}$ would vanish even if $(\xi_P- \xi_S)$ is non-zero. A separation of strong and weak contributions can hence increase the sensitivity to CP violation. As we will discuss next, this is possible through sequential hyperon decays.

\subsubsection{Sequential Decays}
\label{sec:sequence}
Multi-strange and charmed hyperons decay sequentially, \textit{i.e.} into other unstable hyperons. These decay chains give access to the additional decay parameters $\beta$ and $\phi$ \cite{Lee:1957qs}. A formalism for sequentially decaying hyperon-antihyperon pairs from $e^+e^-$ annihilations has been outlined for single-tag measurements in Ref. \cite{Faldt:2017yqt} and for exclusive measurement in Ref. \cite{Perotti:2018wxm}. The latter includes the decay chains of spin 3/2 as well as spin 1/2 hyperons and demonstrates how helicity amplitudes and decay parameters can be extracted from the joint angular distribution of the hyperons and their decay products. 

From the decay parameters $\beta$ and $\phi$, an additional CP test can be constructed:
\begin{equation}
\label{eq:phicp}
    \Delta \phi_{CP} = \frac{\phi_Y+\phi_{\bar{Y}}}{2}. 
\end{equation}

This test has the advantage that it, at first order, only depends on the weak phase difference whereas strong phases do not appear:

\begin{equation}
\label{eq:phicpphase}
    \Delta \phi_{CP} \approx \frac{\left< \alpha_Y \right>(\xi_P-\xi_S)_{LO}}{\sqrt{1-\left<\alpha_Y \right>^2}},
\end{equation}
where 
 \begin{equation}
     \left<\alpha_Y \right> = \frac{\alpha_Y-\alpha_{\bar{Y}}}{2}.
 \end{equation}

This separation of the weak phase difference increases the sensitivity to CP violation by at least an order of magnitude \cite{Donoghue:1986hh}.

\subsubsection{Projected Decay Distributions}

In hyperon beam experiments, where a high-energy hyperon is produced in $pp$ or $pA$ interactions, one can integrate out most angles in the multi-step hyperon decay chain. In other words, one obtains a projection of the angular distribution on the final state baryon angle. This allows for the extraction of the product $\alpha_{Y1}\alpha_{Y2}$ from the final state baryon in the rest frame of the intermediate hyperon $Y_2$:
\begin{equation}
\label{eq:alphaprod}
    \frac{dn}{d\cos\theta_{B}} = \frac{1}{2}(1 + \alpha_{Y1}\alpha_{Y2}\cos\theta_{B}).
\end{equation}

The CP odd quantity $A_{CP}^{\rm \it eff}$ can then be defined

\begin{equation}
\label{eq:acpeff}
    A_{CP}^{\it eff} = \frac{\alpha_{Y1}\alpha_{Y2}+\alpha_{\bar{Y1}}\alpha_{\bar{Y1}}}{\alpha_{Y1}\alpha_{Y2}-\alpha_{\bar{Y1}}\alpha_{\bar{Y2}}} \approx A_{CP}^{Y1} + A_{CP}^{Y2}.
\end{equation}

The advantage of $A_{CP}^{\it eff}$ is its simplicity, but it is also less sensitive to CP violation: It is possible that non-zero CP violations of hyperon $Y_1$ and $Y_2$ have opposite signs and therefore cancel out completely or partly.

\subsection{Radiative Hyperon Decays}
Radiative decays of hyperons, \textit{i.e.} $Y_1 \to Y_2 \gamma$, are either purely electromagnetic (if $Y_1$ and $Y_2$ have the same flavour and hence do not involve transitions between quark generations) or weak (if $Y_1$ and $Y_2$ have different flavour and hence involve transitions such as $s \to u$). Radiative decays into nucleons, \textit{i.e.} $Y \to N\gamma$ are always weak. Electromagnetic decays, \textit{e.g.} $\Sigma^0 \to \Lambda \gamma$, is expected to be parity-conserving, given that i) the $\Sigma^0\Lambda\gamma$ coupling is related to the $nn\gamma$ coupling by SU(3) flavour symmetry and ii) the upper limit of the neutron EDM is found to be extremely small \cite{PhysRevLett.97.131801}. Hence, the decay asymmetry $\alpha_{EM}$, defined by the real part of the product of the electric and magnetic transition dipole moments, should be very close to zero. The $\alpha_{EM}$ is experimentally accessible through the angle between the photon from the $\Sigma^0 \to \Lambda \gamma$ decay and the proton from the subsequent $\Lambda \to p \pi^-$ decay. In Ref. \cite{nair}, it was shown that the asymmetries

\begin{equation}
\label{eq:acpem}
    A^{EM}_{CP} = \alpha^{EM}_{Y}+\alpha^{EM}_{\bar{Y}}
\end{equation}
indicates strong CP violation, while 
\begin{equation}
\label{eq:acem}
    A^{EM}_{C} = \alpha^{EM}_{Y}-\alpha^{EM}_{\bar{Y}}
\end{equation}
indicates C violation. Predictions from the SM of  $A^{EM}_{CP}$ are of the order $10^{-14}$, \textit{i.e.} far below any foreseeable experimental resolution. However, this means that even tiny signals would indicate physics beyond the SM.

Weak radiative decays, involving transitions between quark generations $s \to d\gamma$, can be parameterized in a similar way as the weak hadronic decays, \textit{i.e.} in terms of a decay parameter $\alpha_{\gamma}$. However, since the photon pseudoscalar meson spin are different, the polarization transfer has a different sign. This means that for unpolarized hyperons, the longitudinal polarization of the daughter baryon is $-\alpha_{\gamma}$ in contrast to $+\alpha_Y$ for a hadronic decay, recalling Fig.~\ref{fig:alphabetaphi}. Weak radiative decays contain additional complexity due to the involvement of strong, possibly parity violating processes. According to Hara's theorem \cite{Hara1964}, parity violating amplitudes are predicted to vanish in the limit of SU(3) flavour symmetry, under the assumption that the weak interaction is CP conserving and only includes left-handed currents. This implies that the weak radiative decay asymmetry $\alpha_{\gamma}$ would be zero. This is in contrast to SU(3) breaking where we instead have

\begin{equation}
    \alpha_{\gamma}=\frac{m_{s}^{2}-m_{d}^{2}}{m_{s}^{2}+m_{d}^{2}},
\end{equation}

$m_{s}$ and $m_{d}$ being the mass of the strange and down quark, respectively. Constituent quark masses ($m_{s}\sim 500$~MeV and $m_{d}\sim330$~MeV) yields $\alpha_{\gamma}$ around +0.4 and +0.5. This is similar to the $b\to s\gamma$ transition, where one replaces $m_{s}$ by $m_{b}$. Since $m_{b}>>m_{d}$, the asymmetry parameter will then be close to +1. The $b\to s\gamma$ transition has put constraints on physics beyond the SM that remove the chirality suppression. In a similar way, it is possible to investigate beyond SM effects in $s\to d\gamma$.
%Experimental measurements of photon polarization support that $\alpha_{gamma}=1$ in $b\to s\gamma$ transition. However, it was a surprise when several experiments reported a large negative value of the asymmetry parameter in the $s\to d\gamma$ transition for decays such as $\Sigma^{+}\to p\gamma$, $\Xi^{0}\to\Lambda\gamma$. 
\color{black}

\subsection{Semi-leptonic Hyperon Decays}

Semi-leptonic decays of SU(3) octet hyperons can be parameterized in terms of four weak form factors, which in the limit of zero momentum transfer reduces to three: the vector $g_V(q^2)$, the axial vector $g_A(q^2)$ and the induced tensor $f_2(q^2)$ where $q^2$ is the momentum transfer squared from the mother to the daughter hyperon~\cite{Cabibbo:2003cu,Weinberg:1958ut}. Since the momentum transfer in semi-leptonic hyperon decays is small compared to the baryon masses ($q^2 \to 0$), a systematic expansion can be performed in terms of the symmetry breaking parameter $\delta$, defined by the mass difference of the mother and daughter hyperon \cite{Garcia:1985xz}: $\delta = (M_{Y1}-M_{Y2})/M_{Y1}$. This simplifies significantly the dependence of the decay rate on the form factors. The ratio between the vector and axial vector form factors can be written 
\begin{equation}
\label{eq:gav}
    g_{AV} = \left|\frac{g_A(0)}{g_V(0)}\right|\cdot e^{i\phi_{AV}}.
\end{equation}
The phase $\phi_{AV}$ quantifies the presence of a triple correlation term in the transition probability and a non-zero value would imply violation of time-reversal invariance \cite{ParticleDataGroup:2022pth}. The phase is accessible for polarized mother and/or daughter hyperons from the joint angular distribution of the decay \cite{Frampton:1971sj}. The form factors from different channels are connected through the approximate SU(3) flavour symmetry.

The semi-leptonic hyperon decays from hyperons produced in $e^+e^-$ annihilations can be analysed analogous to hadronic hyperon decays (see Fig.~\ref{fig:decayplanes}) where the scalar particle is replaced by the off-shell gauge boson $W^-$ decaying into an electron-neutrino pair. The study of joint angular distributions typically require large data samples ($N > 1000$). Since semi-leptonic hyperon decays are typically rare compared to hadronic ones, it is challenging to collect the samples necessary for polarization studies. However, also for the modest samples it is possible to gain valuable insights, for example by constructing CP tests from the semi-leptonic decay rates of hyperons and antihyperons: 
\begin{equation}
\label{eq:acpslw1}
    A_{CP}^{\rm \it SLW} = \frac{\Gamma(Y_1 \to Y_2 l^-\bar{\nu}_l)}{\Gamma(\bar{Y}_1 \to \bar{Y}_2 l^+\nu_l)} 
    \end{equation}
    
    \begin{center}
       or 
    \end{center}
    
    \begin{equation}
\label{eq:acpslw2}
    A_{CP}^{\rm \it SLW} = \frac{BR(Y_1 \to Y_2 l^-\bar{\nu}_l)-BR(\bar{Y}_1 \to \bar{Y}_2 l^+\nu_l)}{BR(Y_1 \to Y_2 l^-\bar{\nu}_l)+BR(\bar{Y}_1 \to \bar{Y}_2 l^+\nu_l)}.
\end{equation}
Furthermore, it is straight-forward to construct a test on lepton universality from the ratio
\begin{equation}
\label{eq:rmu}
   R_{\mu e} = \frac{\Gamma(Y_1 \to Y_2 \mu^-\bar{\nu}_{\mu})}{\Gamma(Y_1 \to Y_2 e^-\bar{\nu}_{e})}
\end{equation}
of different lepton decay channels.

\section{Experimental status}

\subsection{Hyperon Structure}

Electromagnetic properties of hadrons have been studied in electron-hadron scattering ever since the 1950s \cite{hofstaedter} which has led to important insights on the structure and size of the proton \cite{Punjabi:2015bba, Pacetti:2014jai, Pacetti:2018wwk,Lin:2021xrc} and to some extent also the neutron \cite{BESIII:2021tbq, Atac:2021wqj}. Due to experimental challenges, corresponding progress for hyperons is yet to come: so far, the only existing charge radius measurement is for the $\Sigma^-$ hyperon \cite{SELEX:2001fbx}. However, the advent of modern, high-intensity $e^+e^-$ colliders, has ignited a veritable revolution for hyperon structure measurements. 

\subsubsection{Effective form factor}

The BaBar experiment carried out a pioneering measurement of the $\Lambda$ and $\Sigma^0$ form factors and the $\Sigma^0\Lambda$ transition form factor by exploiting Initial State Radiation (ISR) on a 230~fb$^{-1}$ sample collected at the bottomium resonance $\Upsilon(4S)$ \cite{BaBar:2007fsu}. Though the resulting hyperon-antihyperon samples were small, they were sufficient to extract the energy dependence of the cross section and the corresponding effective form factor. The CLEO-c experiment studied $\Lambda$, $\Sigma$, $\Xi$ and $\Omega$ production in the $e^+e^- \to Y\bar{Y}$ reaction \cite{Dobbs:2014ifa,Dobbs:2017hyd} at energies corresponding to masses of the charmonium resonances $\Psi(3686)$, $\Psi(3770)$ and $\Psi(4170)$. The results indicated the importance of diquark correlations \cite{Anselmino:1992vg}. However, the interpretation in terms of electromagnetic form factors rely on the assumption that the $Y\bar{Y}$ pair production is dominated by one-photon exchange ($e^+e^- \to \gamma^* \to Y\bar{Y}$) with a negligible contribution from charmonium decay ($e^+e^- \to \Psi \to Y\bar{Y}$). This is in contrast to a recent measurement from the BESIII collaboration \cite{BESIII:2021ccp}, that shows that the branching fraction of the decay $\Psi(3770) \to \Lambda\bar{\Lambda}$ is at least an order of magnitude larger than the prediction used in the CLEO-c papers \cite{Dobbs:2014ifa,Dobbs:2017hyd}. Furthermore, the BESIII results indicate that also other charmonium resonances have a significant impact on the $Y\bar{Y}$ production. This, however, implies that hyperon-antihyperon production in the vicinity of vector charmonium resonances can be utilized for getting deeper insights into the properties of these resonances \cite{Qian:2021neg}.

\begin{figure}[h!]
\begin{center}
%\vspace{-4mm}
\includegraphics[width=0.49\textwidth]{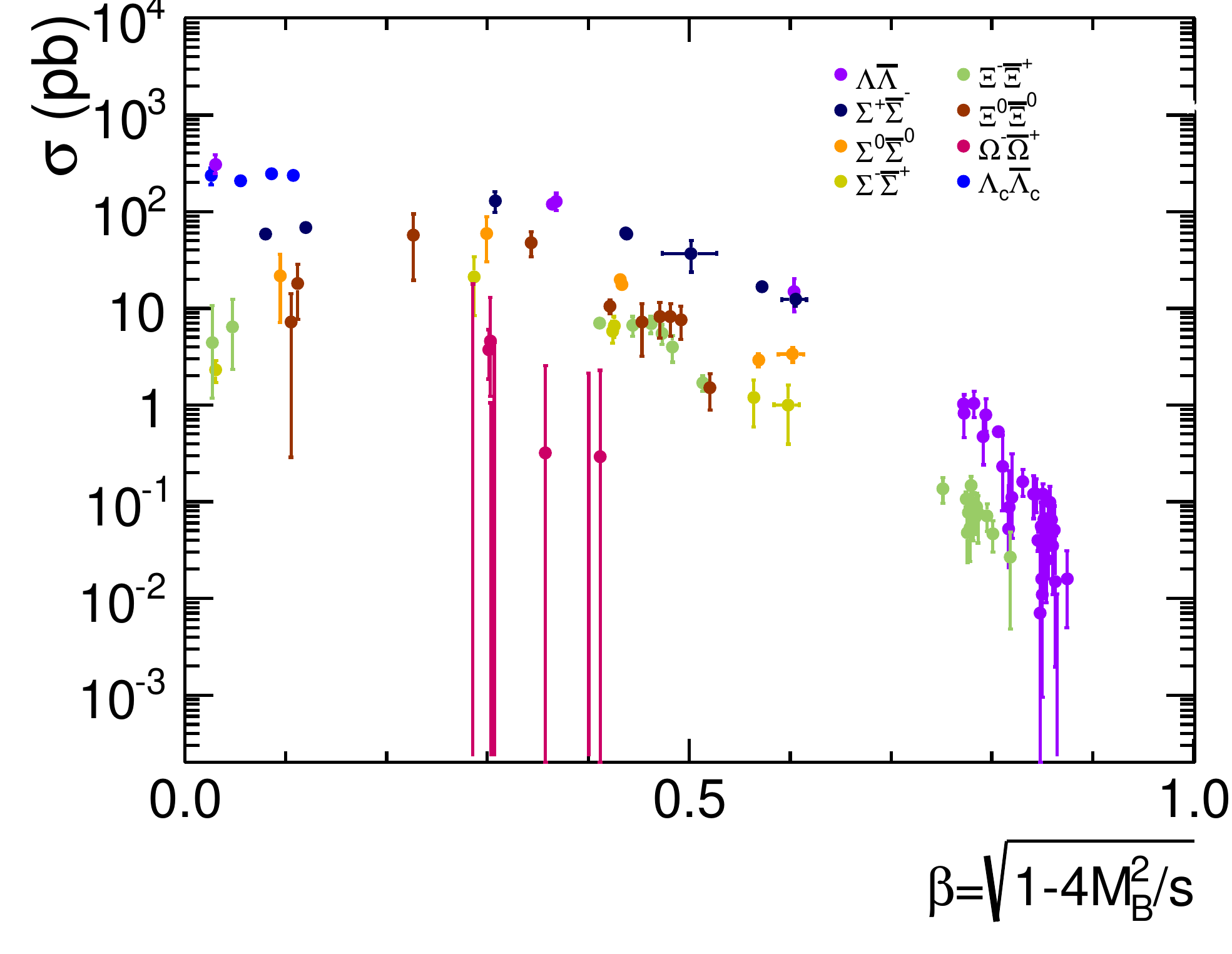}
\includegraphics[width=0.49\textwidth]{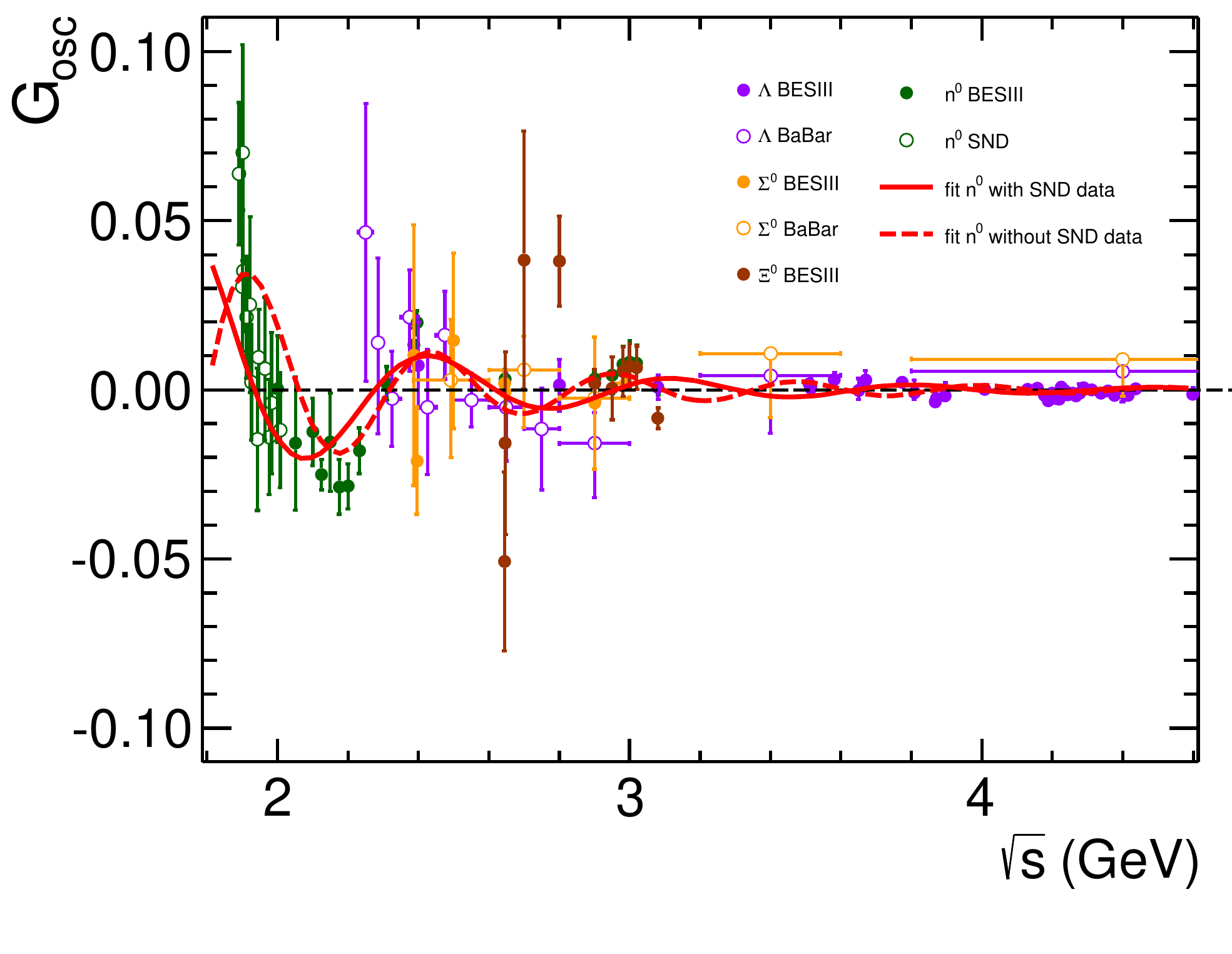}
%\vspace{-7mm}
\end{center}
\caption{Left: The $e^+e^- \to Y\bar{Y}$ cross sections, measured by the BESIII experiment for different ground-state hyperons, as a function of the velocity $\beta$. Right: The effective form factor, after subtracting the dipole term, for neutral baryons studied by BaBar, BESIII and SND experiments, as a function of the square root of the $e^+e^-$ CMS energy. The solid curve represents the fit to BESIII neutron data and the dashed a ft to BESIII and SND data.}
\label{fig:hyptau}
\end{figure}

Dedicated energy scan campaigns by BESIII have enabled increased precision in effective form factor measurements for the full spin $1/2$ octet. A low-energy scan near the $\Lambda\bar{\Lambda}$ threshold revealed an unexpected enhancement, resembling a cusp or a peak \cite{BESIII:2017hyw}. Vector meson dominance, in particular with strong influence by the $\phi(2170)$, has been suggested as an explanation \cite{Cao:2018kos,Li:2021lvs}. Pioneering measurements have also been achieved for $\Sigma^{\pm}$ \cite{BESIII:2020uqk}, $\Sigma^0$ \cite{BESIII:2021rkn}, $\Xi^-$ \cite{BESIII:2019cuv,BESIII:2020ktn} and $\Xi^0$ \cite{BESIII:2021aer}. The production cross sections of these hyperons in the $e^+e^- \to Y\bar{Y}$ process from all BESIII measurements, are summarized in the left panel of Figure~\ref{fig:hyptau}. These results allow for a systematic analysis of the $G_{\rm eff}$ behaviour, similarly to what has recently been done for protons \cite{BaBar:2013ves, BESIII:2021rqk,BESIII:2019hdp} and neutrons \cite{BESIII:2021tbq}. It has been found \cite{Bianconi:2015vva,Bianconi:2015owa} that for nucleons, the effective form factor can be parameterised according to
\begin{equation}
    G_{\rm \it eff}(q^2) = G_D(q^2) + G_{\rm osc}(q^2),
\end{equation}
where $G_D(q^2)$ is a dipole function and $G_{\rm osc}(q^2)$ describes oscillations that are periodic in either $q^2$ or the relative momentum between the nucleon and antinucleon, $p$. The oscillations of the proton and neutron form factors have different amplitude and phase, but the same period \cite{BESIII:2021tbq} and have been attributed to nucleon-antinucleon rescattering into an intermediate hadron-antihadron state. In a recent paper \cite{Dai:2021yqr}, the $G_{\rm \it eff}$ was analysed for neutral octet baryons $n, \Lambda, \Sigma^0$ and $\Xi^0$ using all available data. The study indicated a universal behaviour where $G_{\rm osc}(q^2)$ can be parameterized with a common phase and amplitude, as indicated with a red dashed line in the right panel of Figure\ref{fig:hyptau}. However, their conclusion is not supported by new low-$q^2$ neutron form factor from the SND experiment \cite{SND:2022wdb}. Fitting the oscillation model to both BESIII and SND neutron data results in a significantly lower oscillation frequency than that of the proton. The question remains if the observed oscillations in baryon form factors follow a pattern that can be explained within \textit{e.g.} SU(3) symmetry and hence have a common origin, or is rather accidental.

The structure of the charm baryon $\Lambda_c^+$ has been studied by the Belle experiment using the ISR method \cite{Belle:2008xmh} and by BESIII in an energy scan \cite{BESIII:2017kqg}. The Belle data revealed a peaking cross section near threshold, in contrast to BESIII where a plateau was seen. However, it should be pointed out that the differences between Belle and BESIII are within the experimental uncertainties. Attempts have been made to understand the threshold behaviour of the cross section in terms of vector charmonium resonances \cite{Dai:2017fwx,Song:2022yfr} and final state interactions \cite{Milstein:2022bfg}. 

The cross section of $\Omega^-$ production in $e^+e^- \to \Omega^- \bar{\Omega}^+$ has been studied in a recent analysis of energy scan data from BESIII \cite{BESIII:2022kzc}. Having spin 3/2, the $\Omega^-$ effective form factor is interpreted as a linear combination of the electric, magnetic, quadrupole and octupole form factors \cite{Ramalho:2020laj}. The energy dependence is, within the uncertainties, consistent with the prediction from perturbative QCD and from Ref. \cite{Ramalho:2020laj}.
 
\subsubsection{Separation of Electric and Magnetic Form Factors}

\begin{figure}[h!]
\begin{center}
%\vspace{-4mm}
\includegraphics[width=0.618\textwidth]{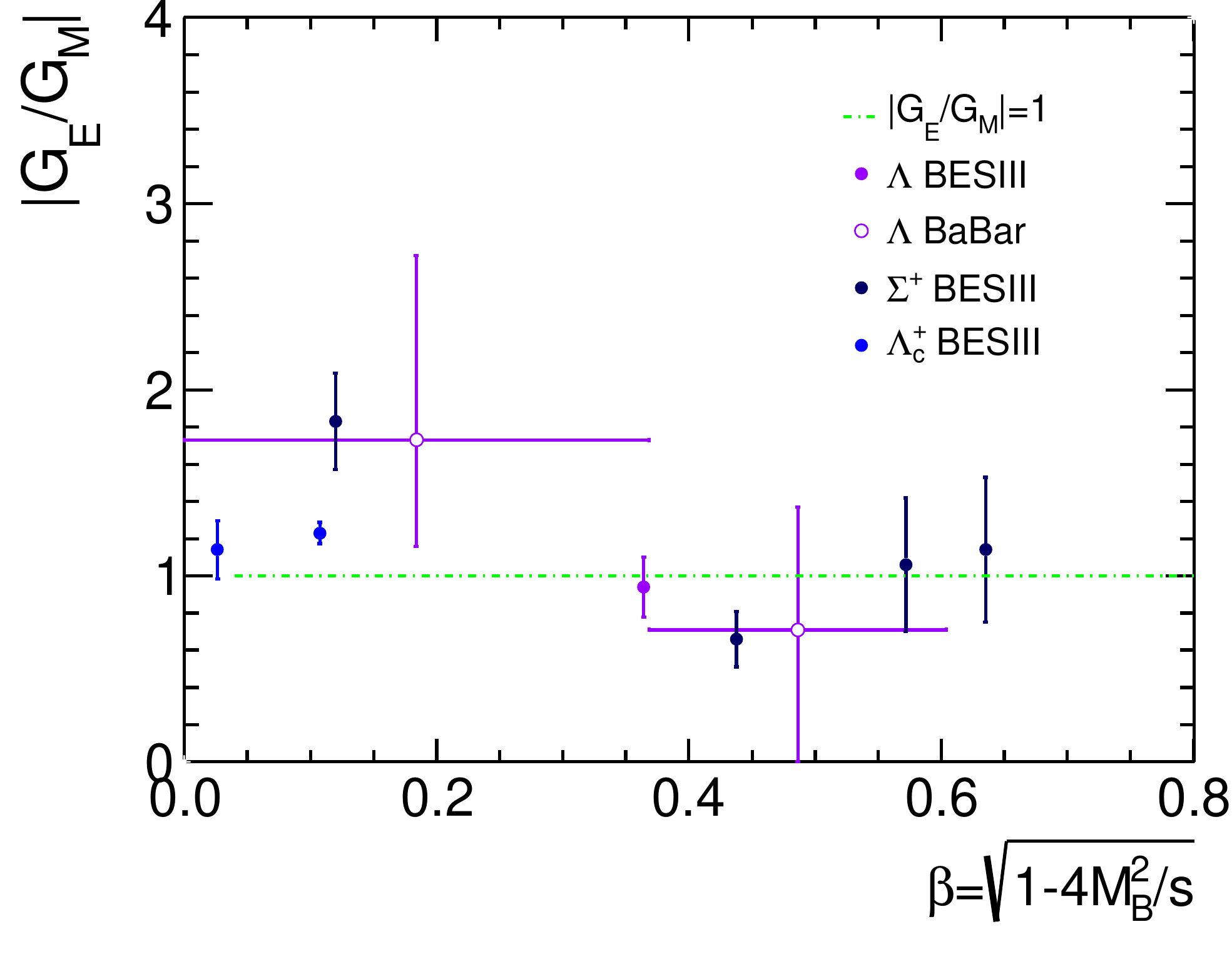}
\end{center}
\caption{ The $|G_{E}/G_{M}|$ of $\Lambda$, $\Sigma^+$ and $\Lambda_c^+$ hyperons as a function of the velocity $\beta$. The filled dots represent data from BESIII and the open dots BaBar.}
\label{fig:hypR}
\end{figure}

The energy scan campaigns carried out by BESIII included particularly large data samples at 2.396~GeV, 2.64~GeV, 2.9~GeV, 4.57~GeV and 4.59~GeV. Using either double-tag ($e^+e^- \to Y\bar{Y}$) or single-tag ($e^+e^- \to Y\bar{X}$ or $e^+e^- \to \bar{Y}X$, where $X$ refers to an unmeasured hyperon) selection, sufficiently many events were found to reconstruct the hyperon scattering angle distribution. This enabled a separation of the electric and the magnetic form factors for $\Lambda$ \cite{BESIII:2019nep}, $\Sigma^+$ \cite{BESIII:2020uqk} and $\Lambda_c^+$ \cite{BESIII:2017kqg}, as summarized in Figure~\ref{fig:hypR}. For $\Lambda$ and $\Lambda_c^+$, the $G_E/G_M$ ratio was found to be close to unity near threshold, whereas for the $\Sigma^+$, it was found to be significantly larger near threshold while approaching unity at larger energies. 

The large sample at 2.396 GeV, in combination with the relatively large $e^+e^- \to \Lambda\bar{\Lambda}$ cross section at this energy, resulted in more than 500 exclusively reconstructed $\Lambda \to p\pi^-, \bar{\Lambda} \to \bar{p}\pi^+$ events. This gave access to the joint angular distribution of the production and decays from which the relative phase $\Delta\Phi$ between $G_E$ amd $G_M$ could be extracted. Hence, the time-like form factors could be unambiguously determined for the first time for any baryon -- a complete snapshot from its time evolution \cite{BESIII:2019nep}. This measurement gave rise to theory activity world-wide \cite{Haidenbauer:2020wyp,Yang:2019mzq,Husek:2019wmt,Junker:2019vvy,Ramalho:2019koj,Bianconi:2022yjq,Lin:2022baj,Dai:2021yqr,Qian:2021neg}. In particular, a new method was outlined which allows for the extraction of the charge radius squared $\left< r^2 \right>$ from polarization measurements \cite{Mangoni:2021qmd}. Combining the $R$ measurement from BESIII and BaBar with the $\Delta\Phi$ measurement from BESIII, a first calculation of $\left< r^2 \right>$ was carried out. However, since $\Delta\Phi$ is only measured at one energy, the result turned out to be ambiguous and six different solutions were presented. Phase measurements in a broader energy range will be helpful in pinpointing the charge radius squared for the first time for a neutral hyperon.

\subsubsection{Hyperon Production at Vector Charmonia}

The large data samples collected by BESIII at charmonium resonances such as $J/\Psi$, $\Psi(3686)$ and $\Psi(3770)$ allow for a very precise determination of the psionic form factor phase $\Delta\Phi_{\Psi}$ and the angular distribution parameter $\alpha_{\Psi}$, as shown in Table \ref{tab:psionic}. It is found that $\Delta\Phi_{\Psi}$ varies substantially between resonances and hyperon types. Summarising the results, we note that
\begin{itemize}
    \item The $\Sigma^+$ phase is negative at the $J/\Psi$ resonance \cite{BESIII:2020fqg} while positive for $\Lambda$ \cite{BESIII:2018cnd} and $\Xi^-$ \cite{BESIII:2021ypr}.
    \item The $\Sigma^+$ phase changes sign when going from the $J/\Psi$ to the $\Psi(3686)$ \cite{BESIII:2020fqg}.
    \item The phase of $\Xi^-$ is significantly larger at $J/\Psi$ \cite{BESIII:2021ypr} compared to at $\Psi(3686)$ \cite{BESIII:2022lsz}.
    \item The angular distribution parameter $\alpha_{\Psi}$ of the $\Psi(3686) \to \Sigma\bar{\Sigma}$ decay is significantly larger for $\Sigma^-$ compared to $\Sigma^+$ and $\Sigma^0$ \cite{BESIII:2022ulr}. 
\end{itemize}

The interpretation of psionic form factors is still a matter of discussion, but the possibility to perform precision measurements calls for a theory framework in which the relation between psionic form factors and charmonium decay is outlined. Hopefully, this could provide a helpful tool to understand the properties of vector charmonia and their coupling to baryon-antibaryon states.

%\textcolor{blue}{PLACEHOLDER: Table with psionic phases $\Delta\Phi$ and angular distribution parameters $\alpha_{\Psi}$ for spin 1/2 hyperons.}

\begin{table}
  \caption{Properties of the $e^+e^-\to J/\psi\to B\overline B$ decays to the pairs of ground-state octet hyperons. \label{tab:psionic}}
%\begin{ruledtabular}
\begin{tabular}{lllll}
  Final state & ${\cal B}(\times 10^{-4})$& $\alpha_\psi$&$\Delta\Phi$ (rad)& Ref.\\ \hline
$J/\psi\to\Lambda\overline{\Lambda}$&$19.43 \pm 0.03 \pm 0.33$&$\phantom{-}0.4748 \pm 0.0022 \pm 0.0031$&$\phantom{-}0.7521 \pm 0.0042 \pm 0.0066$&\cite{BESIII:2017kqw,BESIII:2022qax}\\
$\psi(3686)\to\Lambda\overline{\Lambda}$&$3.97 \pm 0.02 \pm 0.12$&$\phantom{-}0.82 \pm 0.08 \pm 0.02$&$\phantom{-}$--&\cite{BESIII:2017kqw}\\
$\psi(3770)\to\Lambda\overline{\Lambda}$ & $>0.024$ (90$\%$~CL)& $\phantom{-}0.85^{+0.12}_{-0.20}\pm 0.02$ & $\phantom{-}1.24^{+1.15}_{-0.80} \pm 0.09$ &\cite{BESIII:2021ccp,BESIII:2021cvv}\\
\hline
$J/\psi\to\Sigma^+\overline\Sigma\vphantom{X}^-$&$10.61 \pm 0.04 \pm 0.36 $&$-0.508 \pm 0.006 \pm 0.004$&$-0.270 \pm 0.012 \pm 0.009$ &\cite{BESIII:2021wkr, BESIII:2020fqg}\\
$\psi(3686)\to\Sigma^+\overline\Sigma\vphantom{X}^-$&$2.52\pm 0.04 \pm 0.09 $&$\phantom{-}0.682 \pm 0.030 \pm 0.011$&$\phantom{-}0.379 \pm 0.070 \pm 0.014 $ &\cite{BESIII:2021wkr, BESIII:2020fqg}\\
\hline
$\psi(3686)\to\Lambda\overline\Sigma\vphantom{X}^0$&$0.0160 \pm 0.0031 \pm 0.0059$ & -- & -- & \cite{BESIII:2021mus} \\
\hline
$J/\psi\to\Sigma^-\overline\Sigma\vphantom{X}^+$& -- & -- & -- & --\\
$\psi(3686)\to\Sigma^-\overline\Sigma\vphantom{X}^+$&$2.82\pm 0.04 \pm 0.08$ & $\phantom{-}0.96 \pm 0.09 \pm 0.03$ & -- & \cite{BESIII:2022ulr} \\
\hline
$J/\psi\to\Sigma^0\overline\Sigma^0$&$11.64\pm 0.04 \pm 0.23$&$-0.449 \pm 0.020 \pm 0.008$&--&\cite{BESIII:2017kqw}\\
$\psi(3686)\to\Sigma^0\overline\Sigma^0$&$2.44 \pm 0.03 \pm 0.11$&$\phantom{-}0.71 \pm 0.11 \pm 0.04$&--&\cite{BESIII:2017kqw}\\
\hline
$J/\psi\to\Xi^0\overline\Xi\vphantom{X}^0$&$11.65 \pm 0.04 \pm 0.43$&$\phantom{-}0.66 \pm 0.03 \pm 0.05$&--&\cite{BESIII:2016nix}\\
$\psi(3686)\to\Xi^0\overline\Xi\vphantom{X}^0$&$2.73 \pm 0.03 \pm 0.13$&$\phantom{-}0.65 \pm 0.09 \pm 0.14$&--&\cite{BESIII:2016nix}\\
\hline
$J/\psi\to\Xi^-\overline\Xi\vphantom{X}^+$&$10.40 \pm 0.06 \pm 0.74$&$\phantom{-}0.586 \pm 0.012 \pm 0.010$&$\phantom{-}1.213 \pm 0.046 \pm 0.016$&\cite{BESIII:2016ssr,BESIII:2021ypr}\\
$\psi(3686)\to\Xi^-\overline\Xi\vphantom{X}^+$&$2.78 \pm 0.05 \pm 0.14$&$\phantom{-}0.693 \pm 0.048 \pm 0.049$&$\phantom{-}0.667 \pm 0.111 \pm 0.058$&\cite{BESIII:2016ssr,BESIII:2022lsz}\\
\hline
  \end{tabular}
%\end{ruledtabular}
\end{table}
\subsection{Hadronic Hyperon Decays}

\subsubsection{Single-strange Hyperon Decays}

Searches for CP violation in the hadronic $\Lambda \to p\pi^-$ and $\bar{\Lambda} \to \bar{p}\pi^+$ have been carried out since the mid-1980s. The first was performed at the CERN Intersecting Storage Rings in the reactions $pp \to \Lambda X$ and $\bar{p}p \to \bar{\Lambda}X$ \cite{R608:1985fmh}. From the measured $\alpha_{\Lambda} P_{\Lambda}$ and $\alpha_{\bar{\Lambda}} P_{\bar{\Lambda}}$, where $P$ is the polarization, the quantity $\alpha_{\Lambda} P_{\Lambda} / \alpha_{\bar{\Lambda}} P_{\bar{\Lambda}}$ was defined. Assuming $P_{\Lambda} = P_{\bar{\Lambda}}$, this quantity should be 1 when CP is conserved. However, it is straight-forward to construct a CP test  similar to Eq. \ref{eq:acp}

\begin{equation}
\label{eq:acpp}
    A_{CP}^{\alpha P} = \frac{\alpha_{\Lambda}P_{\Lambda}  + \alpha_{\bar{\Lambda}}P_{\bar{\Lambda}}} {\alpha_{\Lambda}P_{\Lambda}  - \alpha_{\bar{\Lambda}}P_{\bar{\Lambda}}} \approx \frac{\alpha_{\Lambda}  + \alpha_{\bar{\Lambda}}} {\alpha_{\Lambda}  - \alpha_{\bar{\Lambda}}}.
\end{equation}
The approximately $10^4$ events yielded the result $A_{CP} = 0.02 \pm 0.14$. The DM2 collaboration measured the $A_{CP}$ using the $J/\Psi \to \Lambda\bar{\Lambda}$ process, however overlooking the possibility that the $\Lambda$ may be polarized. As a result, terms in the joint angular distribution that are sensitive to $\alpha_{\Lambda}/\alpha_{\bar{\Lambda}}$ differences cancel when being integrated over \cite{DM2:1988ppi}. The PS185 collaboration at LEAR, CERN, provided the first essentially model-independent CP test by using the $\bar{p}p \to \bar{\Lambda}\Lambda$ reaction and studying the quantity defined in Eq.~\ref{eq:acpp} as a function of the $\bar{\Lambda}$ scattering angle. This, in combination with the large data sample, $96\cdot10^3$ exclusively reconstructed $\Lambda\bar{\Lambda}$ events, resulted in a significantly improved precision \cite{Barnes:1996si} that remained unchallenged for more than 20 years.

\begin{figure}
    \centering
    \includegraphics[width=1.0\textwidth]{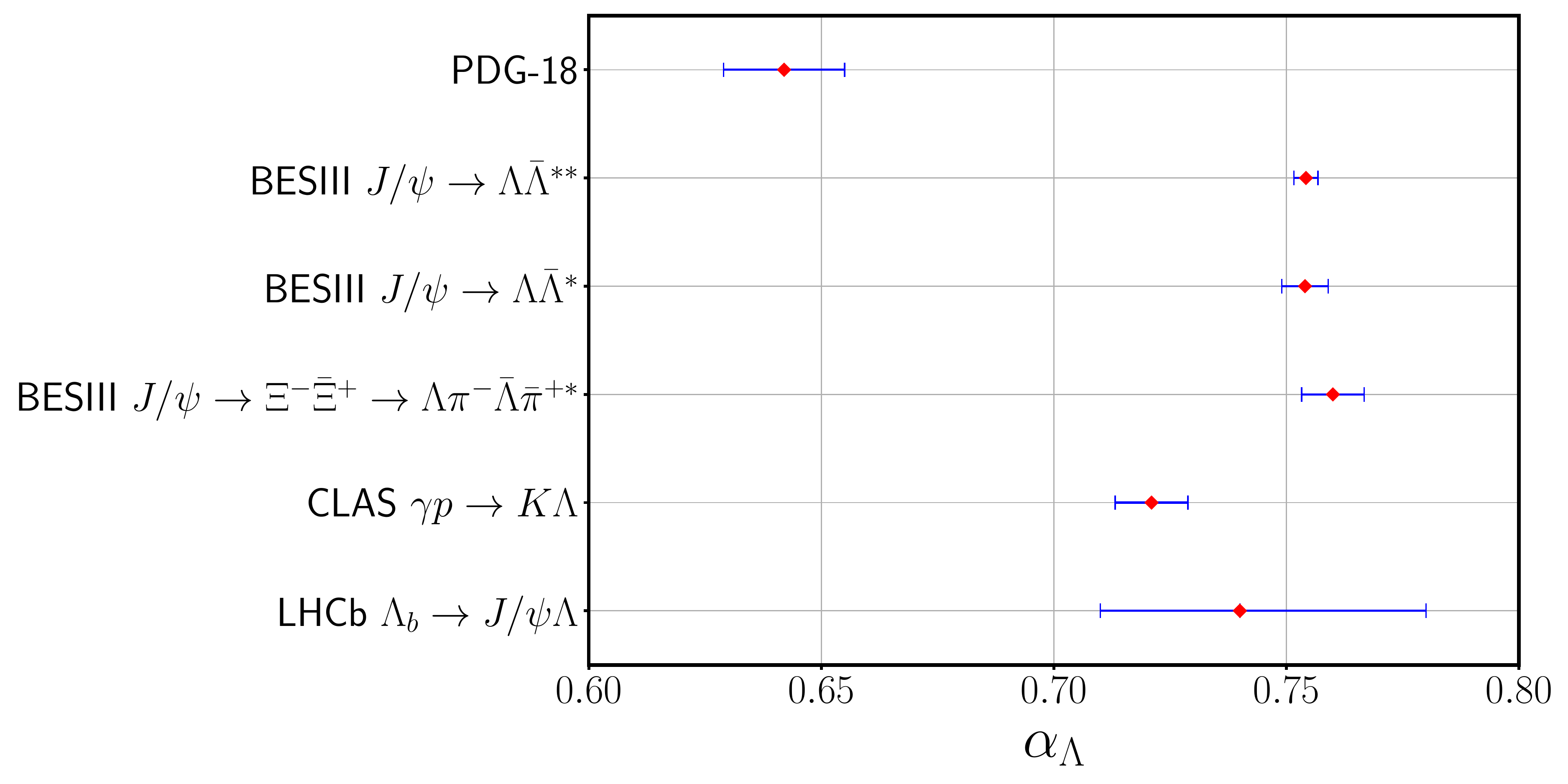}
    \caption{ Asymmetry parameter values $\alpha_{\Lambda}$ from the PDG from 2018 \cite{ParticleDataGroup:2018ovx} and from new measurements by BESIII \cite{BESIII:2018cnd,BESIII:2021ypr, BESIII:2022qax}, CLAS \cite{Ireland:2019uja} and LHCb \cite{LHCb:2020iux}. The PDG-2018 value is based on polarimeter measurements from the 1960s and 1970s. In the new BESIII measurements, the average value $\left< \alpha_{\Lambda}\right> = (\alpha_{\Lambda} - \overline{\alpha}_{\Lambda})/2$ is displayed. The * and ** denote the 2009-2012 and 2009-2019 $J/\psi$ samples, respectively. }
    \label{fig:asymmetry}
\end{figure}

The large $J/\Psi$ samples from the BESIII collaboration -- $1.3\cdot10^{9}$ $J/\Psi$ during 2009 and 2012 and $10^{10} J/\Psi$ during 2018 and 2019 -- started a new era in the search for CP violation in baryon decays. First, the 2009 and 2012 data resulted in $\approx42\cdot10^4$ exclusively reconstructed $\Lambda\bar{\Lambda} \to p\pi^-\bar{p}\pi^+$ events \cite{BESIII:2018cnd} and $\approx88\cdot10^3$ $\Sigma^+\bar{\Sigma}^- \to p\pi^0\bar{p}\pi^0$ events \cite{BESIII:2020fqg}. This enabled the most precise $\Lambda$ CP test so far and the first CP test for $\Sigma^+$. In addition, the asymmetry parameter of the $\bar{\Lambda} \to \bar{n}\pi^0$ was measured for the first time \cite{BESIII:2018cnd}. A remarkable finding in the $\Lambda \to p\pi^-$ decay study was that the asymmetry parameter was found to be 17\% larger than the previous world average, mainly based on proton polarimeter experiments in the 1970s. A re-analysis of CLAS data \cite{Ireland:2019uja} revealed a value more consistent with that obtained by BESIII than the old world average. The Particle Data Group \cite{ParticleDataGroup:2022pth} has now abandoned the old measurements in favour of a world-average based on Refs. \cite{BESIII:2018cnd} and \cite{Ireland:2019uja}. A recent study of the $J/\Psi \to \Xi^-\bar{\Xi}^+ \to \Lambda\pi^-\bar{\Lambda}\pi^+ \to p\pi^-\pi^-\bar{p}\pi^+\pi^+$ provided an independent measurement of $\alpha_{\Lambda}$, $\alpha_{\bar{\Lambda}}$ and $A_{CP}$, consistent with previous results from BESIII. The recent $\alpha_{\Lambda}$ measurements, as well as the previous PDG world average from older experiments, are summarised in Figure~\ref{fig:asymmetry}. 

The precision of the $\alpha_{\Lambda}$ measurement from $\approx73 \cdot10^3$ $\Xi^-\bar{\Xi}^+$ \cite{BESIII:2021ypr} decays is about the same as the one achieved from the almost one order of magnitude larger $J/\Psi \to \Lambda\bar{\Lambda}$ sample \cite{BESIII:2018cnd}. This is because $\Lambda$ hyperons from $\Xi^-$ decays have a larger degree of polarization than $\Lambda$ hyperons from $J/\Psi$ decays. Furthermore, $\Lambda$ hyperons from a sequential decay can be polarized in several directions while those from a decaying $J/\Psi$ can only be polarized along the production plane.

Even more recently, the full $10^{10} J/\Psi$ sample, including $3\cdot10^6$ exclusively reconstructed $\Lambda\bar{\Lambda}$ decays, has been analyzed for the purpose of achieving a high-precision CP test. The $\Lambda \to p\pi^-$ decay asymmetry was found to be consistent with the previous measurement and the precision of $A_{CP}$ is the most precise so far: $0.0025 \pm 0.0046 \pm 0.0012$.

\subsubsection{Double-strange Hyperon Decays}

Sequential decays of multi-strange and charmed hyperons give, in addition to the decay asymmetry $\alpha_Y$, also access to $\beta$ and $\gamma$ or $\phi$ (see section \ref{sec:sequence}). The studies of these parameters are not limited to exclusive, double-tag $Y\bar{Y}$ measurements but can also be studied when it is only possible to reconstruct either the hyperon decay chain or the antihyperon ditto, \textit{i.e.} using the \textit{single-tag} approach.

The world data bank \cite{ParticleDataGroup:2022pth} of the $\Xi^0$ and $\Xi^-$ hyperons is dominated by experiments of polarized or unpolarized secondary hyperon beams. Many of these apply the formalism in Eq.~\ref{eq:alphaprod}. The most precise measurement of the $\alpha_{\Xi^0}\alpha_{\Lambda}$ was achieved by the NA48 experiment at CERN \cite{Batley:2010bp}. The $\phi_{\Xi0}$ parameter was measured with large uncertainties at various experiments \cite{PhysRevD.9.49,PhysRev.179.1262,PhysRev.147.945}.

The E756 collaboration at Fermilab achieved the best precision so far of the $\alpha_{\Xi^-}\alpha_{\Lambda}$ product and also measured the $A_{CP}^{\rm \it eff}$ defined in Eq.~\ref{eq:acpeff} for the first time \cite{E756:2000rge}. Thanks to a dedicated measurement, the HyperCP collaboration improved the precision of $A_{CP}^{\rm \it eff}$ and their measurement is the most precise so far: $(0.0 \pm 5.1 \pm 4.4)\times 10^{-4}$ \cite{HyperCP:2004zvh}. 

By studying the joint angular distribution of polarized $\Xi^-$ hyperons, the E756 \cite{PhysRevLett.91.031601} and HyperCP \cite{HyperCP:2004not} collaborations also measured the $\phi_{\Xi}$ parameter, from which the strong phase difference ($\delta_P - \delta_S$) was extracted. These two measurements implied a significant improvement of the world data bank \cite{PhysRevD.32.2270,Bensinger:1988va,PhysRevD.9.49,Cool:1974df,PhysRevLett.29.1630,PhysRev.179.1262, PhysRev.147.945}. Indeed, the ($\delta_P - \delta_S$) was found to be very small, showing that even a CP violating weak phase would be difficult to detect from $A_{CP}$ using Eq.~\ref{eq:acpphase}. 

A recent measurement from BESIII demonstrates the merits of polarized, entangled and sequentially decaying hyperon-antihyperon pairs \cite{BESIII:2022qax}. The full process $e^+e^- \to \Xi^-\bar{\Xi}^+, \Xi^- \to \Lambda \pi^-, \Lambda \to p \pi^- + c.c.$ was studied using $\approx73\cdot10^3$ exclusively reconstructed, or \textit{double-tagged}, $\Xi^-\bar{\Xi}^+$ pairs identified from a sample of $1.3\cdot10^{9}$ $J/\Psi$. The asymmetries $A_{CP}^{\Xi}$ and  $A_{CP}^{\Lambda}$ were determined separately. From $\Delta \phi_{\Xi}$, the weak phase difference $\xi_P - \xi_S$ has been determined for the first time in a direct measurement. The result, $(\xi_P - \xi_S) = (1.2 \pm 3.4 \pm 0.8)\cdot10^{-2}$ rad, is consistent with CP conservation. It is also noteworthy that the strong phase difference $\delta_P - \delta_S$ was measured with a similar precision as the one achieved by HyperCP \cite{HyperCP:2004not} despite the 2000 times larger data sample from HyperCP. The excellent sensitivity achieved by the BESIII method can be attributed to the fact that the clean, well-known production process and the subsequent decays can be treated in a model-independent way: all angles are measured and no information is integrated out from the differential cross sections. Hence, all quantum correlations between the hyperon and the antihyperons, and their decay products, are taken into account, as well as the polarization transfer in multiple directions from the mother $\Xi^-$ to the daughter $\Lambda$ hyperon. Extending this analysis to the eight times larger data sample collected in 2018 and 2019 will further improve the precision and enable comparison with theory calculations of the strong phase from \textit{e.g.} \cite{PhysRevD.96.016021}. 

\subsubsection{Triple-strange Hyperon Decays}

The triple-strange $\Omega^-$ hyperon should, according to the quark model, have spin 3/2. However, this was not empirically well-established until recently; the first measurements from SPS at CERN \cite{Aachen-Berlin-CERN-Innsbruck-London-Vienna:1977ojz,Birmingham-CERN-Glasgow-MichiganState-Paris:1978nrw} were based on very small data samples and therefore very simplified angular distributions. In a later measurement by BaBar, $\Omega^-$ hyperons were studied in $\Omega_c$ and $\Xi_c$ decays and the conclusions relied on assumptions of the helicity of the mother baryons \cite{BaBar:2006omx}. Recently, the BESIII collaboration applied the model-independent formalism outlined in Ref. \cite{Perotti:2018wxm} on single-tagged $\Omega^- \to \Lambda K^-, \Lambda \to p \pi^-$ and $\bar{\Omega}^+ \to \bar{\Lambda} K^+, \bar{\Lambda} \to \bar{p} \pi^+$ from the $e^+e^- \to \Psi(3686) \to \Omega^-\Omega^+$ reaction. The sample contained $\approx450\cdot10^6$ $\Psi(3686)$ events, including 2507 $\Omega^-$ and 2238 $\bar{\Omega}^+$ candidates. Two hypotheses were tested: $\Omega^-$ having either spin 1/2 or 3/2. The spin 3/2 turned out to be favoured over the spin 1/2 hypothesis by 14$\sigma$ \cite{BESIII:2020lkm}, which is the most precise, model-independent spin test for the $\Omega^-$ hyperon so far. This validates not only the quark model but also the results based upon its validity. In addition, the decay asymmetry parameter $\phi_{\Omega}$ was measured for the first time.

In the past, the decay asymmetries of the decays $\Omega^- \to \Lambda K^-$, $\Omega^- \to \Xi^0 \pi^-$ and $\Omega^- \to \Xi^- \pi^0$ have been studied with a secondary hyperon beam at CERN \cite{BRISTOL-GENEVA-HEIDELBERG-ORSAY-RUTHERFORD-STRASBOURG:1984jku}. More recently, the HyperCP experiment measured the decay asymmetry of $\Omega^- \to \Lambda K^-$ and $\bar{\Omega}^- \to \bar{\Lambda} K^+$ and the CP asymmetry $A_{CP}$. The HyperCP employed a secondary $\Omega$ beam produced in $p Cu$ collisions and since all $\Omega$ hyperons were selected to be collinear with the proton beam, they were unpolarized. This enabled the measurement of the $\alpha_{\Omega}\alpha_{\Lambda}$ and $\alpha_{\bar{\Omega}}\alpha_{\bar{\Lambda}}$ products. By combining these quantities with $\alpha_{\Lambda}$ and $\alpha_{\bar{\Lambda}}$ from other measurements, it was concluded that the $\Omega^- \to \Lambda K^-$ decay conserve parity almost completely. Their measured values of $A_{CP}$ as defined by Eq.~\ref{eq:acp} and the effective asymmetry $A_{CP}^{\rm \it eff}$ by Eq.~\ref{eq:acpeff} were both found to be consistent with CP conservation.

\subsubsection{Charm Hyperon Decays}

Charm hyperons can decay into many different final states and as a consequence, the branching fraction of each channel is small. Therefore, charm hyperon decays are almost exclusively studied using the single-tag approach.

Most measurements of the $\Lambda_c^+$ decay parameters, by \textit{e.g.} ARGUS \cite{ARGUS:1991yzs}, CLEO \cite{CLEO:1993fhs,CLEO:1995qyd} and FOCUS \cite{FOCUS:2005vxq} collaborations only consider the angular distribution of the final state hadron, \textit{i.e.} not the joint decay of the multi-step process. This means that the decay asymmetry $\alpha_{\Lambda_c}$ can be retrieved, but not $\beta$ and $\gamma/\phi$. In addition, the results rely on a precise knowledge of the decay parameter of the hyperon in the second step, for example $\Lambda$ or $\Sigma^+$. Nevertheless, the FOCUS collaboration performed the first CP test for a charmed baryon, by measuring $A_{CP}$ for the $\Lambda_c^+ \to \Lambda\pi^+$ decay to be $0.07 \pm 0.19 \pm 0.24$. 

In a recent BESIII measurement, based on 0.5 fb$^{-1}$ at a $e^+e^-$ CMS energy of 4.6 GeV, we presented a proof-of-concept for measuring the decay parameter $\alpha_{\Lambda_c}$, and for the first time also $\beta_{\Lambda_c}$ and $\gamma_{\Lambda_c}$, of the $\Lambda_c^+$ decay into $\Lambda\pi^+$, $\Sigma^+\pi^0$ and $\Sigma^0\pi^+$ \cite{BESIII:2019odb}, all containing hyperons that decay subsequently. In addition, $\alpha_{\Lambda_c}$ of the $\Lambda_c^+ \to pK_S$ was measured for the first time, though with large uncertainties. Due to limited sample size, it was assumed that $\alpha_{\Lambda_c} = - \alpha_{\bar{\Lambda_c}}$ \textit{i.e.} CP symmetry. The BESIII collaboration has also carried out the first model-independent experimental determination of the $\Lambda_c^+$ spin. The analysis combined several sequential two-body decay chains and found that the spin 1/2 hypothesis  was favoured over the spin 3/2 hypothesis by more than 6$\sigma$ \cite{BESIII:2020kap}. 

The most precise measurements of $\alpha_{\Lambda_c}$ for several two-body decays have been achieved by the Belle experiment in the $e^+e^- \to \Lambda_c^+ X + c.c.$ reaction \cite{Belle:2022uod}. The CP asymmetry $A_{CP}$ has been calculated and though the statistical precision is comparable to that of strange hyperon decays, the total uncertainty is dominated by the systematics. The Belle \cite{Belle:2021crz,Belle:2021zsy} and CLEO \cite{CLEO:2000lsg} experiments have also measured the $\alpha_{\Xi_c}$ of several $\Xi_c^0$ decays.

\subsection{Radiative Hyperon Decays}

From Hara's theorem \cite{Hara:1964zz} and from the breaking of SU(3) symmetry in the quark model, on expects a positive value of the weak radiative decay parameter $\alpha_{\gamma}$ for decays involving $s \to d$ transitions. One example of such a decay is $\Sigma^+ \to p\gamma$. Therefore, it came as a surprise when several experiments reported a large and negative value \cite{PhysRev.188.2077,PhysRevLett.68.3004,PhysRevLett.59.868,MANZ1980217}. Other experiments reported similar values for the $\Xi^0 \to \Lambda \gamma$ and $\Xi^0 \to \Sigma^0 \gamma$ decay \cite{2004251,KTeV:2000dsr,Batley:2010bp}, in disagreement with older measurements \cite{James:1990as,Teige:1989uk}. The discrepancy between theory predictions and experimental results sparked wide interests and various solutions to the puzzle were proposed \cite{Gavela1981417,Zenczykowski:1999vq,Borasoy:1999nt,Zenczykowski:2005cs}. It has been suggested that the $\Lambda\to n\gamma$ can provide further insights \cite{Larson:1993ig,Zenczykowski:2020hmg}: if positive, it indicates a violation of Hara's theorem.

Polarized and entangled hyperon-antihyperon pairs from vector charmonium decays have the advantage that the polarization of the hyperon (antihyperon) can be measured with high precision if one considers the "leg" of the reaction where a hyperon (antihyperon) decays into a two-body hadronic final state. A simultaneous measurement of the other radiative weak decay "leg" of the antihyperon (hyperon) will then also be precise, thanks to constrained production parameters such as the polarization and spin correlations. This has been demonstrated in a recent study by the world-record data sample of $10^{10}$ $J/\Psi$ decays from BESIII. These data contain an appreciable fraction of $J/\Psi \to \Lambda\bar{\Lambda}, \Lambda \to n\gamma, \bar{\Lambda} \to \bar{p}\pi^+ + c.c.$ events that can be exclusively reconstructed. The large transverse polarization of $\Lambda$ \cite{BESIII:2018cnd} enabled the first determination of the decay parameter $\alpha_{\gamma}$ of $\Lambda\to n\gamma$. It was found to be $-0.16\pm0.10_{\rm stat.}\pm0.05_{\rm syst.}$ \cite{BESIII:2022rgl}, indicating that Hara’s theorem still holds for hyperon radiative decays. The results provide essential input to constrain the amplitudes of parity violation contribution in theoretical calculations, and can be used to validate the predictions and constrain the parameters of the various theoretical models \cite{He:1999ik,Tandean:1999mg,Shi:2022dhw}. Recent calculations based on Chiral Perturbation Theory are in good agreement with the BESIII data \cite{Shi:2022dhw}.

\subsection{Semi-leptonic Hyperon Decays}

The small branching ratios and the immense difficulty in detecting neutrinos make the study of semi-leptonic hyperon decays a challenging task. On the one hand, a semi-leptonic decay can be identified also without the neutrino thanks to the special characteristic of a three-body decay compared to the more abundant two-body hadronic decays. On the other hand, the angular analyses from which the physics is revealed, require all other particles in the decay to be accurately reconstructed. Nevertheless, experiments with polarized beams made pioneering measurements of $\Lambda \to p e^-\bar{\nu}_{e}$ in the 1970s \cite{Argonne-Chicago-OhioState-WashingtonUniversity:1971hso,Lindquist:1975fc,Burnett:1976fr,Althoff:1973my,Althoff:1971hxl,Baggett:1972jnv,Canter:1971zj,Maloney:1969xd}, though with relatively modest sample sizes. Later measurements offered larger samples \cite{Wise:1980xx,Bristol-Geneva-Heidelberg-Orsay-Rutherford-Strasbourg:1983jzt,Dworkin:1990dd} and enabled measurements of the axial vector form factor ratio $g_A/g_V$ for $\Lambda$ \cite{ParticleDataGroup:2022pth}. More recently, the KTeV \cite{KTeVE832E799:1999tte,KTeV:2005nvk,KTeV:2001djr} and NA48 \cite{NA48I:2006yat,NA481:2012dtx} collaborations have studied the decay $\Xi^0 \to \Sigma^+e^-\bar{\nu}_{e}$, providing the form factor ratios $g_1(0)/f_1(0)$ and $f_2(0)/f_1(0)$. In addition, the KTeV measured $g_2(0)/f_1(0)$.

Semi-leptonic decays provide an independent way of measuring the CKM matrix element $|V_{us}|$, in addition to kaon and $\tau$ decays. The current world-average of $0.2243\pm0.0008$ is essentially calculated from $K\to\pi l\nu(\gamma)$  and $K\to\mu\nu(\gamma)$ decays \cite{ParticleDataGroup:2022pth}. Combining all available data on semi-leptonic hyperon decays provide comparable precision as for kaon decays and can thereby further improve the precision of the average. The fit of previous measurements of semi-leptonic hyperon decays yields $|V_{us}|=0.2250\pm0.0027$~\cite{Cabibbo:2003cu}. Further theoretical and experimental progress should improve the precision in the semi-leptonic hyperon decays. 

The abundant production of polarized hyperons and antihyperons in the BESIII experiment, in combination with the versatile and near 4$\pi$ detector, enables a new generation of semi-leptonic hyperon decay studies. Recent measurements have been carried out of decays such as $\Lambda\to p\mu^-\bar{\nu}_{\mu}$ \cite{BESIII:2021ynj} and $\Sigma^+\to\Lambda e^+\nu_{e}$~\cite{BESIII:2022lsr}. In the latter study, the CP asymmetry defined in Eq. \ref{eq:acpslw1} was measured for the first time. Furthermore, searches have been performed for the expectedly rare $\Xi^- \to \Xi^0e^-\bar{\nu}_{e}$ \cite{BESIII:2021emv} and $\Xi^0 \to \Sigma^-e^+\nu_{e}$ \cite{BESIII:2022bys} decays. The results from the BESIII semi-leptonic studies are summarised in Table \ref{tab:semileptonic}.

All aforementioned results are obtained by analyzing only the semi-leptonic decay itself, \textit{i.e.} not considering the production process. The ability of the BESIII experiment to reconstruct the complete $Y\bar{Y}$ pair and its subsequent hadronic and semi-leptonic decays, opens up for a new method where the production is taken into account. The method outlined in Ref. \cite{Perotti:2018wxm} can be generalized to describe the joint angular distributions \cite{Batozskaya:2023rek}. It allows for the extraction of spin correlations and polarization of the produced hyperon-antihyperon system which improves the precision of the semi-leptonic decay parameters (Eq.~\ref{eq:gav}).
%\textcolor{blue}{Question: Available material focus on BRs and decay widths, which breaks the narrative a bit compared to the rest of the paper. Maybe a larger focus on angular distributions and polarisation?}
%\textcolor{blue}{Summarise results in a table}

\begin{table}
\label{tab:semileptonic}
  \caption{BRs measurement of semi-leptonic hyperon decays in the $e^+e^-\to J/\psi\to Y\overline Y$ decays at BESIII. The data statistics are $10^{10} J/\psi$ events. \label{tab:SLW}}
%\begin{ruledtabular}
\begin{tabular}{lllllll}
  $Y\bar{Y}$ pair & Final state & ${\cal B}(\times 10^{-4})$& $N^{\mathrm{SLW}}$&$R_{\mu e}$ &$A_{CP}^{SLW}$ &Ref.\\ \hline
$J/\psi\to\Lambda\overline{\Lambda}$&$\Lambda\to p\mu^-\bar{\nu}_{\mu}$&$1.48\pm0.21\pm0.08$&$64\pm9$&$0.178\pm0.028$&&\cite{BESIII:2021ynj}\\
\hline
$J/\psi\to\Xi^-\overline{\Xi}^+$&$\Xi^-\to\Xi^0 e^-\bar{\nu}_e$&$<2.59(90\%)$&$-$&&&\cite{BESIII:2021emv}\\
\hline
$J/\psi\to\Xi^0\overline{\Xi}^0$&$\Xi^0\to\Sigma^- e^+\nu_e$&$<1.6(90\%)$&$-$&&&\cite{BESIII:2022bys}\\
\hline
$J/\psi\to\Sigma^+\overline{\Sigma}^-$&$\Sigma^+\to\Lambda e^+\nu_e$&$29.3\pm7.4\pm1.3$&$16\pm4$&&$1.06\pm0.28$&\cite{BESIII:2022lsr}\\
\hline
  \end{tabular}
%\end{ruledtabular}
\end{table}

A data collection campaign during 2019-2021, where 4.5 fb$^{-1}$ of data were collected between 4.6 GeV and 4.7 GeV, enabled a pioneering study of the differential decay rate and weak form factors of the semi-leptonic charm decay $\Lambda_c^+\to\Lambda e^+\nu_e$ ~\cite{BESIII:2022ysa}. The results were compared to predictions from lattice QCD calculations \cite{Meinel:2016dqj}. By combining the measured branching fraction, the life-time $\tau_{\Lambda_c}$ and the $q^2$-integrated rate predicted by lattice QCD, the matrix element $|V_{cs}|$ was determined to be $|V_{cs}|=0.936\pm0.017_{\mathcal{B}}\pm0.024_{\mathrm{LQCD}}\pm0.007_{\tau_{\Lambda_c}}$. The semi-leptonic $\Lambda_c^+$ decay measurement provides an important consistency test for the SM and a probe for physics beyond the SM, complementary to $D$ meson analyses~\cite{Richman:1995wm,FlavourLatticeAveragingGroupFLAG:2021npn}. 

Searches for violations of lepton flavour universality in hyperon decays have been studied in the numerous hyperon beam experiment in the 1970s~\cite{Ronne:1964zz,Lind:1964zz,Canter:1971ay,Baggett:1972ms}. The most precise measurement was performed in 1972~\cite{Baggett:1972ms} and was reported as a relative  $\Gamma(\Lambda\to p\mu^-\bar{\nu}_{\mu})/\Gamma(\Lambda\to N\pi)=(1.4\pm0.5)\cdot10^{-4}$ based on 14 signal events which were selected from about $6\cdot10^5$ bubble chamber pictures. The current world average of the observable $R_{\mu e}$, defined in Eq.~\ref{eq:rmu}, is $0.189\pm0.041$ for $\Lambda$ \cite{ParticleDataGroup:2022pth}. The large experimental uncertainty is dominated by the $\mathcal{B}(\Lambda\to p\mu^-\bar{\nu}_{\mu})$. Recently, the BESIII collaboration~\cite{BESIII:2021ynj} reported a value of $0.178\pm0.028$ based on 31 signal events obtained from the $10^{10}$ $J/\psi$ sample. All results so far are consistent with the SM prediction, obtained at next-to-leading order, of $0.153\pm0.008$~\cite{Chang:2014iba}. A more accurate measurement of $R_{\mu e}$ will provide a more stringent test of lepton flavour universality.

\section{Discussion}

%\input{PatrikDiscussion.tex}

%\textcolor{red}{NOTE: The discussion is for discussing what we have learned specifically from recent results, mainly from BESIII}

The abundant production of polarized and entangled hyperon-antihyperon pairs in the BESIII experiment has paved the way for a new generation of measurements on structure and symmetry. In the following, we will summarise the recent findings and formulate the next steps in the future roadmap of hyperon physics.

What is the origin of the oscillatory features of the form factors (effective form factor, $G_E/G_M$ ratio and magnetic form factor) observed for protons, nucleons and possibly also hyperons? The first comparison between protons \cite{BESIII:2019hdp} and neutrons \cite{BESIII:2021tbq} indicated oscillations with a common frequency, a behaviour that possibly carries over to strange hyperons. However, a recent measurement by the SND \cite{SND:2022wdb} instead reveals a smaller frequency for neutrons which raises the question of whether the observed oscillatory patterns are only similar by accident or if they indeed have a common origin. Mapping the form factors over a broad energy range for the complete spin 1/2 octet would be very illuminating and can help find an answer to this question. Data for such measurements can either be obtained in dedicated energy scans with the BESIII experiment or by utilizing the technique of initial state radiation (ISR) \cite{Bonneau:1971mk} on data collected at charmonium or bottomium resonances with the BESIII (charmonium) and Belle II (bottomium) experiments.

How well can we understand the structure of hyperons on the femtometer scale? Can we extract space-like quantities using time-like measurements? A recent publication \cite{Mangoni:2021qmd} outlines a method to quantitatively relate the experimentally accessible time-like region to space-like properties, such as the charge radius, from the energy dependence of the form factor phase. So far, the only existing conclusive, off-resonance form factor phase measurement is for the $\Lambda$ hyperon \cite{BESIII:2019nep}. This corresponds to a "snapshot" of the time evolution of the hyperon-antihyperon pair. The energy dependence of the phase yields a sequence of such snapshots, from which the space-like structure can be revealed. Corresponding measurements for the full SU(3) octet or even to charm hyperons, can reveal patterns from which we will gain a deeper understanding of the complex dynamics that bind quarks into hadrons.

At $e^+e^-$ collision energies that coincide with vector charmonia such as $J/\Psi$, $\Psi(3686)$ and $\Psi(3770)$, large data samples are available from the BESIII experiment. This has enabled very precise measurements of the so-called psionic form factors \cite{Faldt:2017kgy}. However, it is yet an open question how to interpret these form factors and what they can reveal about the decaying charmonium resonance. In particular, the close connection between the form factors and polarization could be helpful in understanding the role of spin in charmonium decay processes. However, in order to optimize the output from the high-precision data, the measured parameters need a quantitative foundation in the form of a phenomenological model. An increased exchange between the experimental collaborations and the theory community can lead to new ideas and help exploiting the richness of the hyperon-antihyperon decay distributions.

The abundant production of hyperon-antihyperon pairs whose decays constitute a hunting-ground for CP violation, complementary to that of \textit{e.g.} the $\epsilon$ and $\epsilon'$ parameters in kaon decays. This is particularly true in the search for BSM physics, since many such models which involve an enhanced parity-even gluon dipole operator that does not contribute to $\epsilon'$. Parameters such as $A_{CP}$ and $(\xi_P - \xi_S)$ are sensitive to both parity-odd and parity-even operators and can hence be non-zero even if $\epsilon'$ is vanishingly small. For several observables, the CP violating signal is expected to be up to two orders of magnitude larger for certain classes of models beyond the SM compared to the SM \cite{He:1991pf,He:1995na,He:1999bv,Tandean:2002vy,HE20221840}. More precise CP tests will hence provide constraints for BSM models and guidance for future theoretical efforts. In addition, increased precision in the measurement of \textit{e.g.} strong phases will be helpful in theory calculations of weak phases. One example is the $\Sigma$ decay in Ref. \cite{Tandean:2002vy} for which the theory uncertainty of the weak phase is $\approx$200\%. New, precise measurements of the decay amplitudes of various hyperons will increase the precision of the predictions from theory.  But how precise measurements are possible?

\begin{figure}
    \centering
    \includegraphics[width=1.\textwidth]{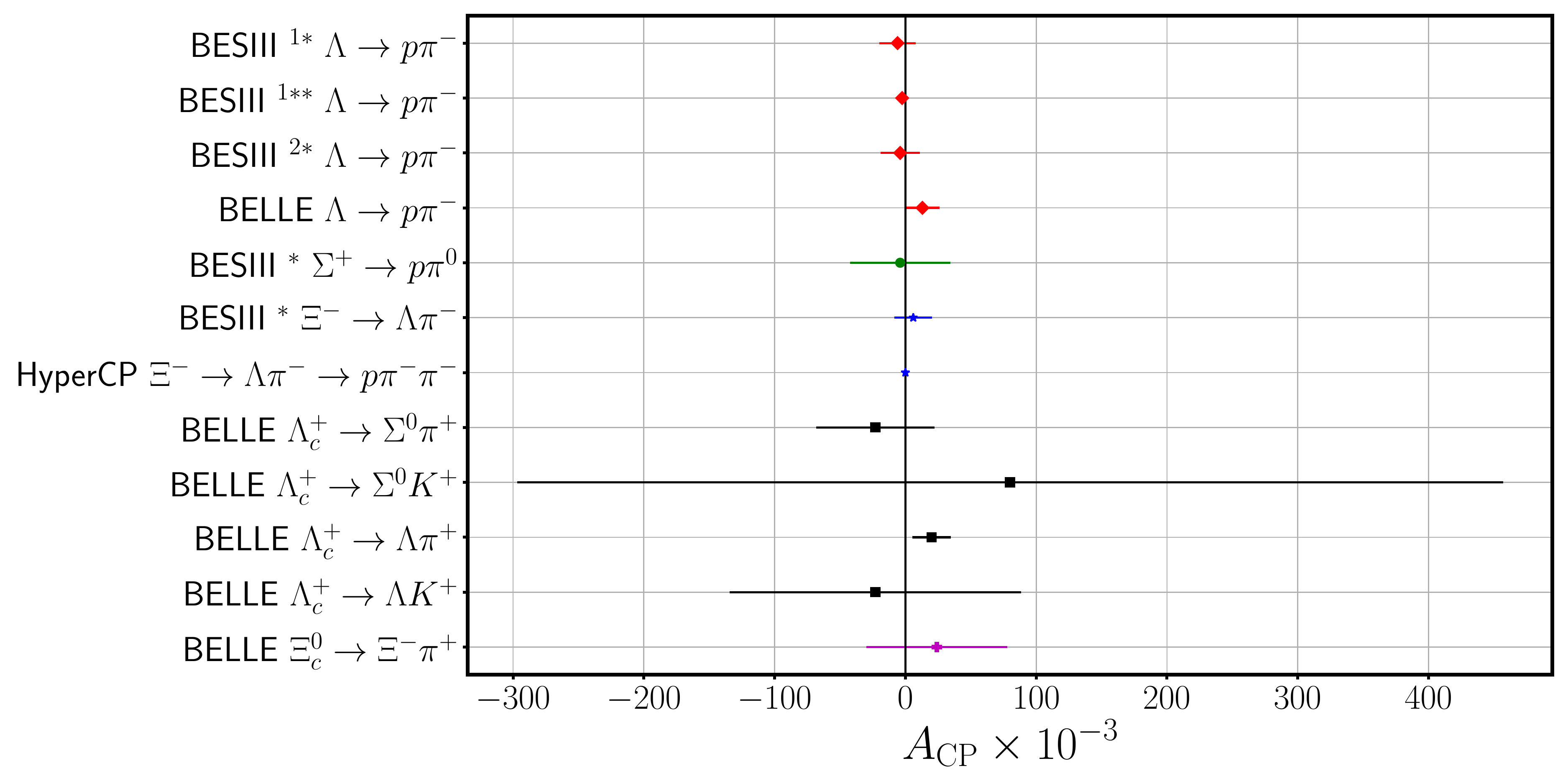}
    \caption{ CP symmetry tests from Refs. \cite{BESIII:2018cnd, BESIII:2021wkr, BESIII:2021ypr, BESIII:2022qax, HyperCP:2004zvh, Belle:2021crz, Belle:2022uod}. The $J/\psi\to \Lambda\overline{\Lambda}$ (1) and $J/\psi\to \Xi^{-}\overline{\Xi}^{+}\to \Lambda \pi^{-}\overline{\Lambda}\pi^{+}$ (2) from BESIII have both been used to determine $A_{CP}$ for $\Lambda\to p \pi^{-}$. The asterisks * and ** denote the 2009-2012 and 2009-2019 $J/\psi$ samples, respectively. In the sequential decay from HyperCP, the effective asymmetry $A_{CP}^{\rm \it eff}$, defined in Eq. \ref{eq:acpeff}, was measured. }
    \label{fig:ACP}
\end{figure}

Recent measurements from the BESIII experiment, exploiting the polarization and spin correlations, provide increased precision for a given sample size \cite{BESIII:2021ypr} compared to techniques based on polarized beams of secondary hyperons \cite{HyperCP:2004not}. In particular, sequentially decaying multi-strange and charmed baryons offer a way to separate strong and weak contributions to the decay amplitude, giving clean access to CP-violating signals. In Figure~\ref{fig:ACP}, the CP asymmetry $A_{CP}$ is shown for a number of hyperon decays. Though many of these are very precise, it is not sufficient to test the aforementioned theoretical predictions. In fact, the precision required is out of reach with currently available data samples. The ongoing campaign of collecting 50~ab$^{-1}$ of data with Belle II, as well as the planned next-generation facilities PANDA and STCF will extend the current data bank with at least two orders of magnitude. Experiments with polarized electron-positron beams can further increase the precision \cite{Salone:2022lpt}.

With current facilities such as BESIII and Belle II, methods based on entangled hyperon-antihyperon pairs are in principle only possible to apply to strange hyperons. Charmed baryons decay into a large number of final states, all with relatively small branching fractions (at percent level or smaller). The recent single-tag measurements by Belle II of $\Lambda_c^+$ and $\Xi_c^0$ decays have in some cases a precision approaching that of strange hyperons. However, a limiting factor is the systematic uncertainties. This may be partly attributed to the method in which the polarization of the initial and intermediate baryons in the chain $Y_1 \to Y_2 M_1 \to Y_3 M_1 M_2 \to ...$ are integrated out. The ongoing Belle II campaign, where up to 50~ab$^{-1}$ of data will be collected, opens up for methods in which the richness of the joint sequential angular distribution can be exploited to its fullest \cite{Faldt:2017yqt}. This can potentially improve the precision of charmed baryon decay parameters.

The large transverse polarization of hyperons such as $\Lambda$, $\Sigma^+$ and $\Xi^-$ produced in vector charmonium decays, open up for precision studies of decay parameters in weak, radiative $Y_1 \to Y_2 \gamma$ decays. These decays provide a tool to understand the elementary process $s \to d \gamma$ in a way that is complementary to radiative kaon decays, which in turn serve as guidance in the search for physics beyond the SM \cite{He:1999ik,Tandean:1999mg}.

Due to the small branching fractions, the semi-leptonic data samples are small. As a consequence, the focus of semi-leptonic studies being put on observables based on decay widths rather than angular distributions and polarization. Large samples from ongoing campaigns or future facilities can be a game-changer in this regard, and lead to substantial advances.

\section{Future Prospects}

In the following, we will discuss the existing facilities BESIII and Belle II, as well as the planned PANDA and STCF, where polarized and entangled hyperon-antihyperon pairs can be studied.

\subsection{Existing Facilities}

\subsubsection{BESIII at BEPCII}

%\textcolor{blue}{Figure with data points of integrated luminosities (acquired and planned) as a function of CMS energy?}

\begin{figure}[h!]
\begin{center}
%\vspace{-4mm}
\includegraphics[width=0.49\textwidth]{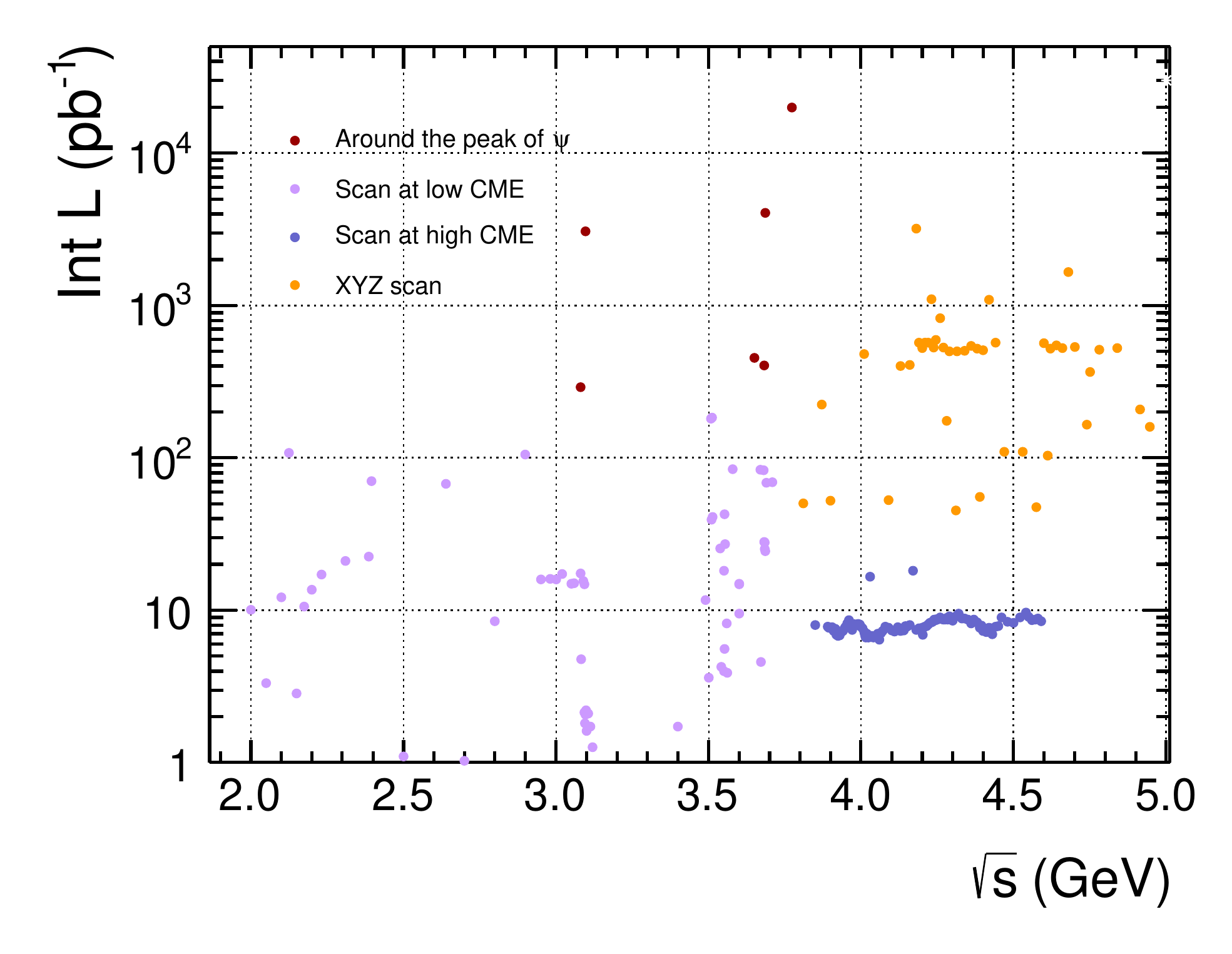}
\includegraphics[width=0.49\textwidth]{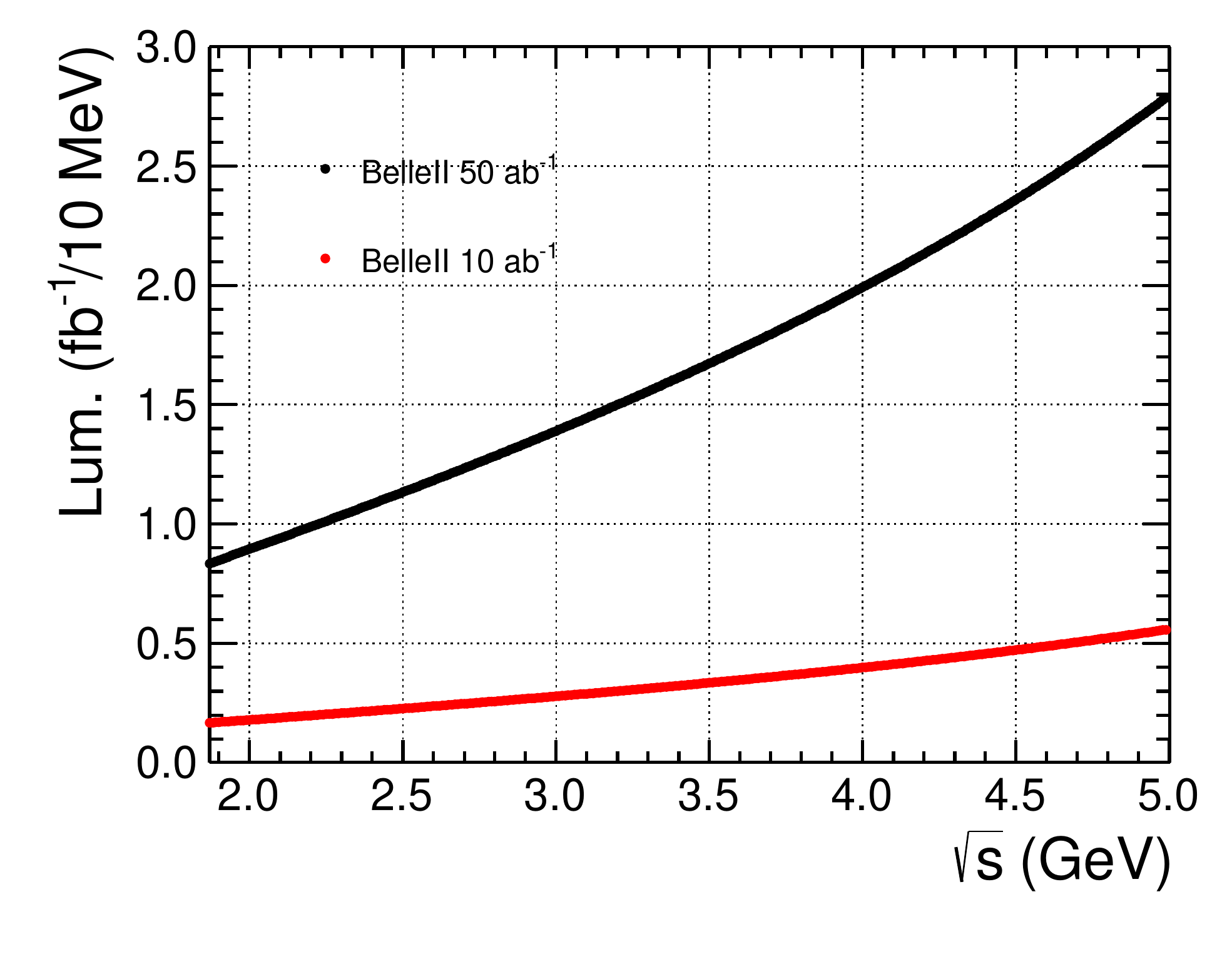}
\end{center}
\caption{Left: The integrated luminosity collected with BESIII at BEPCII till 2024, as a function of the CMS energy $\sqrt(s)$. Right: The effective integrated luminosity with ISR with a reduced $\sqrt(s)$ between 1.87 and 5.0 GeV with the Belle~II experiment. }
\label{fig:besbelle}
\end{figure}

Data from the energy scan performed by BESIII in 2014-2015 have resulted in several pioneering hyperon structure measurements so far \cite{BESIII:2019nep,BESIII:2021rkn,BESIII:2020uqk}, but the story does not end here. Several analyses of these data are still ongoing and the results will be published soon. These studies include $\Lambda$ and $\Sigma^+$ form factor phase measurements at several energies. 

A campaign to collect 20~fb$^{-1}$ data at the $\Psi(3770)$, would enable precision studies of all the SU(3) octet hyperons using the ISR technique. Furthermore, a dedicated campaign has been carried out between CMS energies of 4.6 GeV and 4.95 GeV, with the purpose of a precision measurement of the $q^2$ dependence of the $\Lambda_c^+$ form factors. 

At the time of this review, the most precise $\Lambda$ CP test so far has emerged from the world-record data sample of $10^{10}$ events. However, data from the same sample are being analysed for high-precision CP tests in $\Sigma^0$, $\Sigma^+$, $\Xi^-$ and $\Xi^0$ decays. Since most of these hyperons decay into an intermediate $\Lambda$, each of these studies will provide an independent measurements of the $\alpha_{\Lambda}$ parameter. In addition, this and future BESIII samples will enable searches for rare hyperon decays with unprecedented precision. In addition to the aforementioned radiative and semi-leptonic decays, it is possible to search for rare or forbidden processes, for example those involving flavour changing neutral currents. 

The physics potential of BESIII will reach a new level with a planned upgrade of the BEPCII collider to a maximum beam energy of 2.8~GeV. This corresponds to a CMS energy of 5.6~GeV where the peak luminosity will be increased by a factor of three compared to what is possible at CMS energies below 4.0~GeV. As a result, BESIII will offer a great opportunity to study the production and structure of various charmed baryons, as well as their decay properties. In the left panel of Figure \ref{fig:besbelle}, the integrated luminosities previous and planned data campaigns are summarized.

\subsubsection{Belle II at SuperKEKB}

 The goal of the ongoing campaign with the Belle II experiment at SuperKEKB in Japan \cite{Ohnishi:2013fma} is to collect in total 50~ab$^{-1}$ of data in the bottomium region, with an intermediate milestone of 10~ab$^{-1}$. This enables high-precision structure studies of strange and charmed hyperons through the ISR technique: although the radiator function in the ISR cross section results in a reduced effective luminosity (see right panel of Figure~\ref{fig:besbelle}). The huge integrated luminosity will eventually outperform other off-resonance samples from other $e^+e^-$ colliders.
 
The Belle II and its predecessor Belle have, with the data already on tape, performed pioneering measurements of charmed baryon decay asymmetry parameters, as well as CP tests. New data have been and are being collected by Belle II and can provide better statistical precision. In addition, larger samples enable studies of more complete decay chains, exploiting the richness of the angular distributions. This could reduce the systematic uncertainties which currently dominate the total uncertainty. 

\subsection{Future Facilities}
The methods developed for hyperon-antihyperon pairs provide sensitive CP tests, but are currently limited by statistical precision. Improving the uncertainties by several orders of magnitude is not feasible with currently operating facilities world-wide, but requires next-generation experiments where polarized and entangled/quantum-correlated hyperon-antihyperon pairs are produced in abundance. The PANDA experiment at FAIR, Germany and the proposed STCF at USTC, China are two excellent candidates which will be presented in the following.

\subsubsection{PANDA}

The future experiment antiProton ANihilation at DArmstadt (PANDA) will produce hyperon-antihyperon pairs in abundance, thanks to the expected large production cross section in $\bar{p}p$ annihilations and the large luminosities enabled by the stored antiproton beam. The latter will be delivered from the High Energy Storage Ring (HESR) within a momentum range from 1.5 GeV/$c$ up to 15 GeV/$c$ and impinge on a hydrogen cluster or pellet target. During the first phase of operation, \textit{Phase One} \cite{PANDA:2021ozp}, the HESR will provide a luminosity of $\approx 10^{31}$~cm$^2$s$^{-1}$ whilst the design luminosity of $\approx 2\cdot10^{32}$~cm$^2$s$^{-1}$ will be achieved during Phase Three. Since PANDA will cover nearly $4\pi$ of the solid angle, the prerequisites for exclusive reconstruction of hyperon-antihyperon pairs and their subsequent decays are well fulfilled. This has been demonstrated by simulation studies in for example Refs. \cite{PANDA:2020zwv,PANDA:2021ozp}, summarised in Table \ref{tab:hypprod}. We conclude that PANDA, already during the first phase of operation, will be able to collect samples several orders of magnitude larger than what is currently possible with BESIII. What does this mean in terms of the precision of the CP asymmetries? 

\begin{table}
\centering
\caption{Results from simulation studies of the various production reactions of ground state hyperons \cite{PANDA:2020zwv,PANDA:2021ozp}.}
\label{tab:hypprod}       % Give a unique label
% For LaTeX tables you can use
\begin{tabular}{llllll}
\hline
Reaction & $p_{\overline{p}}$ (GeV/$c$)  & S/B & Events / day, Phase One & Events / day, full luminosity
\\\hline
$\overline{p}p \rightarrow \overline{\Lambda}\Lambda$ & 1.64 &  114 & $3.8\cdot 10^6$ & $7.6\cdot 10^7$ \\\hline

$\overline{p}p \rightarrow \overline{\Sigma}^0\Lambda$ & 6.0 & 21 & $4.3\cdot 10^5$  & $4.1\cdot 10^6$ \\\hline

$\overline{p}p \rightarrow \overline{\Xi}^+\Xi^-$ & 4.6 & 274 & $2.6\cdot 10^4$  & $5.2\cdot 10^5$ \\\hline
\end{tabular}
% Or use
%\vspace*{1cm}  % with the correct table height
\end{table}

In $\bar{p}p$ annihilations, many partial waves contribute to the production. This is in contrast to the $e^+e^-$ case, where the spin and parity of the initial state is almost exclusively $J^P = 1^-$. Hence, in $e^+e^-$ production, all production parameters such as polarization and spin correlations are well-defined functions of the hyperon scattering angle $\theta_Y$, governed by two global parameters $\alpha_{\Psi}$ and $\Delta\Phi$ that depend on the collision energy only. Hyperon-antihyperon pairs produced in $\bar{p}p \to \bar{Y}Y$ reactions are quantum correlated rather than entangled since several possibilities exist for the initial spin-parity state. Nevertheless, in the case of spin 1/2 hyperons, the reaction can be parameterized in terms of five independent production parameters, each with an unknown $\theta_Y$ dependence. Decay parameters and CP asymmetries, therefore, need to be extracted from global fits. However, as demonstrated in Ref. \cite{Barnes:1996si}, the precision for a given sample size is about the same as in $e^+e^-$ experiments \cite{BESIII:2018cnd}. The future CP tests by PANDA really have the potential to significantly improve the precision of the world data and put SM and BSM predictions to the test.

\subsubsection{STCF}

The Super Tau-Charm facility~(STCF) proposed in China is a next-generation electron-positron collider that will operate within a CMS energy range from 2.0 to 7.0~GeV~\cite{Peng:2022loi}. The design luminosity is $0.5\cdot10^{35}$~cm$^{-2}$s$^{-1}$ or higher at the optimal CMS energy of 4.0~GeV. With this luminosity, it will be possible to collect 1~ab$^{-1}$ of data per year. At the $J/\Psi$ mass, this corresponds to $3.4\cdot 10^{12}$ $J/\Psi$ events per year, given a production cross-section of 3400~nb and a beam energy spread that equals that of BESIII. This opens up for high-precision CP tests in decays of hyperon-antihyperon pairs. Realistic projections of the precision can be obtained by scaling the statistical uncertainty achieved with BESIII with at least two orders of magnitude larger data sets expected with the STCF. The results are presented in Table~\ref{tab:ALCP} and are discussed in detail in Ref.~\cite{Salone:2022lpt}.

\begin{table}
\centering
\caption{Expected statistical uncertainty for the CP asymmetries $A_{CP}^{\Lambda}$, $A_{CP}^{\Xi^-}$ and $B_{CP}^{\Xi^-}$ at BESIII and the proposed STCF. BESIII has already published measurements on $A_{CP}^{\Lambda}$~\cite{BESIII:2018cnd,BESIII:2022qax} and $A_{CP}^{\Xi^-}$~\cite{BESIII:2021ypr}, the former with the full $10^{10} J/\Psi$ data sample. The projected uncertainties for future measurements at BESIII and STCF are obtained by scaling the statistical uncertainties with the number of events. The $\ast$ and $\ast\ast$ indicate the combination of BESIII measurements~\cite{BESIII:2018cnd,BESIII:2022qax} and projected precision, respectively.} 
\label{tab:ALCP}
% For LaTeX tables you can use
\begin{tabular}{lllll}
\hline
&$N_{J/\psi}$&$\sigma(A_{CP}^{\Lambda})$&$\sigma(A_{CP}^{\Xi^-})$&$\sigma(B_{CP}^{\Xi^-})$\\
    \hline
    BESIII &$1.3\cdot10^{9}$&$1.0\cdot10^{-2\ast}$&$1.3\cdot10^{-2}$&$3.5\cdot10^{-2}$\\
    BESIII &$1.0\cdot10^{10\ast\ast}$ &$3.6\cdot10^{-3}$&$4.8\cdot10^{-3}$&$1.3\cdot10^{-2}$\\
    STCF &$3.4\cdot10^{12\ast\ast}$&$2.0\cdot10^{-4}$&$2.6\cdot10^{-4}$&$6.8\cdot10^{-4}$\\ 
    \hline
  \end{tabular}\\
% Or use
%\vspace*{1cm}  % with the correct table height
\end{table}

Future upgrades may include even higher luminosities, a longitudinal polarization of the electron beam and a reduced beam energy spread. It is estimated that an electron beam polarization of about $80\%$ at $J/\psi$ energies can be obtained without reducing the beam current~\cite{Koop:talk}. This polarization is much larger than the observed natural transverse polarization of hyperons and antihyperons (about 30\% at its maximum) from $J/\psi$ decays and will increase the sensitivity of most CP tests in hyperon decays ~\cite{Salone:2022lpt}.

The beam energy spread can be reduced by a collision scheme where electrons (positrons) with higher momenta collide with positrons (electrons) with lower momenta~\cite{Renieri:proc,Avdienko:1983mee,Telnov:2020rxp}, similar to the Belle II experiment at SuperKEKB \cite{Ohnishi:2013fma}. Reducing the CMS energy spread compared to what is currently possible with BEPC-II can further increase the data quality. In particular, if the spread approaches the natural width of the $J/\psi$, $\rm{\Gamma}=90$~keV, the number of the $J/\psi$ events produced with a given integrated luminosity will increase. Furthermore, the off-resonance background can be reduced which would improve the sensitivity to rare processes. 

\section{Summary}

Polarized and entangled hyperon-antihyperon pairs offer a common platform to study a wide range of topics including precision aspects of the strong interaction, tests of fundamental symmetries and flavour physics. Recent measurements from BESIII employ newly developed methods that benefit from the possibility to precisely parameterize the complicated decay chains, including particle-antiparticle correlations. This has resulted in pioneering measurements of the hyperon structure at the femtometer scale, some of the most precise CP tests for baryons so far, and studies of rare hyperon decays. These methods can be extended and applied to larger data samples from BESIII and Belle II, but also from future particle-antiparticle collider experiments such as PANDA and STCF. The results can lead to a more complete understanding of the Standard Model with all its features, including currently challenging aspects such as the strong interaction in the confinement domain. Furthermore, future hyperon-antihyperon studies have the potential to reveal physics beyond the Standard Model. 

\section*{Acknowledgement}

The authors are grateful to the Knut and Alice Wallenberg Foundation, Contract No. 2016.0157 and 2021.0299 (Sweden), the Swedish Research Council, Contracts No. 2019-04594 and 2021-04567 (Sweden), The Swedish Foundation for International Cooperation in Research and Higher Education, Contract No. CH2018-7756, the Olle Engkvist Foundation, Contract No. 200-0605 (Sweden), Lundström-Åman Foundation (Sweden), the National Natural Science Foundation, Contract Nos. 11935018, 12122509 (China), the CAS President’s International Fellowship Initiative, Grant No. 2021PM0014 (China), Polish National Science Centre, Grant 2019/35/O/ST2/02907 (Poland), the Double First-Class university project foundation of USTC (China). Finally, we would like to thank Prof. Hai-Bo Li for valuable input and encouragement during the writing process.

%Polarized and entangled pairs of hyperons and antihyperons provide excellent tools to study aspects of the Standard Model, such as the strong and weak interaction, but also to search for physics beyond it. In particular, spin properties that are accessible in the rich decay distribution, provide information on structure and fundamental symmetries that neither stable protons or spinless mesons can reveal. The electron-positron collider experiment BESIII has entered the precision mode thanks to dedicated energy scans as well as a world-record sample of $10^{10}$ collected $J/\Psi$. 

%\bibliographystyle{unsrtnat}
\bibliographystyle{ieeetr}
\bibliography{references}  %%% Uncomment this line and comment out the ``thebibliography'' section below to use the external .bib file (using bibtex) .



% 	\newblock Fast classification of handwritten on-line arabic characters.% 	\newblock In {\em Soft Computing and Pattern Recognition (SoCPaR), 2014 6th


% \end{thebibliography}

\end{document}